\documentclass[10pt,prd,twocolumn,preprint,nofootinbib]{revtex4}

\usepackage{epsfig}
\usepackage{amsmath,amsfonts,amssymb}
\usepackage{t1enc}
\usepackage{verbatim}
\usepackage{float}
\usepackage{morefloats}

\providecommand{\openone}{\leavevmode\hbox{\small1\kern-3.8pt\normalsize1}}

\usepackage{booktabs}
\usepackage[Q=yes,pverb-linebreak=no]{examplep}

\newcommand{\be}{\begin{equation}}
\newcommand{\ee}{\end{equation}}

\def\mgamc{{{\tt \small MG5\_aMC}}}

\begin{document}

\title{\boldmath Pseudoscalar couplings in $t \bar{t} H$ production at NLO+NLL accuracy \unboldmath}

\author{
Alessandro Broggio$^{1}$,
Andrea Ferroglia$^{2,3}$,
Miguel C. N. Fiolhais$^{4,5}$,
Antonio Onofre$^{6}$
\\[3mm]
{\footnotesize {\it 
$^1$Physik Department T31, Technische Universit\"at M\"unchen, James Franck-Stra{\ss}e 1, D-85748 Garching, Germany \\
$^2$Physics Department, New York City College of Technology, The City University of New York, 300 Jay Street, Brooklyn, NY 11201 USA \\
$^3$The Graduate School and University Center, The City University of New York, 365 Fifth Avenue, New York, NY 10016  USA \\
$^4$ Science Department, Borough of Manhattan Community College, City University of New York, \\ 199 Chambers St, New York, NY 10007, USA \\
$^5$ LIP, Departamento de F\'{\i}sica, Universidade de Coimbra, 3004-516 Coimbra, Portugal\\
$^6$ LIP, Departamento de F\'{\i}sica, Universidade do Minho, 4710-057 Braga, Portugal\\
}}
}

\preprint{TUM-HEP-1086/17}

\begin{abstract}
We study the production of a Higgs boson in association to a top-antitop pair at the Large Hadron Collider. We show how precise predictions for the differential distributions with respect to the transverse momentum of the Higgs boson, to the invariant mass of the top-antitop-Higgs system and to the invariant mass of the top-antitop pair can provide useful information on the possible presence of a pseudoscalar component in the coupling of the top quark with the Higgs boson. We evaluate the production of a top-antitop pair and a Higgs boson to next-to-leading order in fixed order perturbation theory and we carry out the resummation of soft emission corrections to next-to-leading-logarithmic accuracy for the LHC operating at a center of mass energy of $13$~TeV. We discuss how the shape of these distributions can be employed experimentally, making a physics case for the kinematic reconstruction of dilepton channels.
\end{abstract}

\maketitle

\section{Introduction}
\label{sec:intro}

At the Large Hadron Collider (LHC), the study of the associated production of a top-quark pair and a Higgs boson offers the 
unique opportunity of obtaining direct information on the Yukawa coupling of the top quark. Since in the Standard Model (SM) of 
particle physics the top quark is the particle which is predicted to have the strongest coupling with the Higgs boson, 
the study of this Yukawa coupling can prove important in the understanding of the electroweak symmetry breaking mechanism. 
In order to fully exploit the potential of the LHC in the study of this process, precise measurements, involving ideally the full kinematic reconstruction of the top pair and Higgs boson momenta, should be matched by (at least) equally precise theoretical predictions for measured observable quantities.
For this reason, within the SM, higher order corrections to this process have been a subject of investigation for several 
years. Next-to-leading-order (NLO) QCD corrections were at first evaluated in \cite{Beenakker:2001rj,Beenakker:2002nc,Reina:2001bc,Reina:2001sf,Dawson:2002tg,Dawson:2003zu}. Subsequently, these corrections were evaluated again in the process of developing and testing new  tools for automated calculation of  
NLO corrections  \cite{Frederix:2011zi,Garzelli:2011vp}. The electroweak corrections to this process were studied in 
\cite{Yu:2014cka,Frixione:2014qaa,Frixione:2015zaa,Hartanto:2015uka}. The associated production of a top pair and a Higgs boson, including the decay of the top quark and off-shell effects was evaluated to NLO in \cite{Denner:2015yca,Denner:2016wet}.

The resummation of potentially large effects due to the emission of soft gluons in the final state was  studied in \cite{Broggio:2015lya,Broggio:2016lfj,Kulesza:2015vda,Kulesza:2016vnq,Kulesza:2017ukk}. In~\cite{Broggio:2016lfj} a parton level Monte Carlo code was developed in order to evaluate the soft emission corrections to $t \bar{t} H$ production to next-to-next-to-leading logarithmic (NNLL) accuracy. 
The NNLL corrections were then matched to NLO calculations obtained by employing \verb!MadGraph5_aMC@NLO! \cite{Alwall:2014hca} (which we will indicate with {\mgamc} in the rest of this paper). In this way, within the SM, it was possible to obtain the NLO+NNLL prediction for the $t \bar{t} H$ total cross section as well as for several differential distributions which depend on the four momenta of the final state top quark, top antiquark and Higgs boson.

The cross section of the associated production of the Higgs boson with a top-quark pair at the LHC is  of a few hundred femtobarn. Since this process has a  huge background contamination stemming from $pp \rightarrow t\bar{t}+\textrm{jets}$ and other processes, it is easy to see that the associated $t \bar{t} H$ productions is extremely difficult to measure at the LHC. The ATLAS and CMS analysis teams have already obtained a combined best-fit value for the signal strength of $\mu = \sigma/\sigma_{\text{SM}} = 2.3^{+0.7}_{-0.6}$~\cite{Aad:2016zqi,Khachatryan:2014qaa,Khachatryan:2016vau}. The expected increase of integrated luminosity at the LHC, and further progress in the experimental techniques might make the the observation possible at run II.

It is of course also crucial to try to understand which kind of information about possible non-SM couplings of the top 
quark to the Higgs boson can become accessible when this process will be detected at the LHC. 
In particular, it is interesting to see if the Higgs boson couples to the top quark not only with a scalar coupling, but also through a pseudoscalar one. 
A comprehensive study of this aspect can  be found in \cite{Demartin:2014fia}, where the CP properties of the top-quark Yukawa interaction are studied in the context of Higgs production in gluon fusion or in association with top quarks. In that work, NLO QCD preditions for $t \bar{t} H$ production within an Effective Field Theory approach of \cite{Artoisenet:2013puc} are presented.

This works investigates the possible presence of a pseudoscalar component in the top-quark Yukawa coupling from a different perspective; in particular  
the purpose of this paper is twofold:
\begin{itemize}
\item[\emph{a)}] We reconsider the role that the shape of some differential distributions can have in the study of a possible pseudoscalar component in the coupling of the Higgs boson to the top-quark, taking into account the possibility of a full experimental reconstruction of the top, antitop and Higgs boson momenta.
\item[\emph{b)}] We show that it is relatively straightforward to obtain precise, beyond NLO predictions for differential distributions which are sensitive to a pseudoscalar component in the top quark Yukawa coupling.
\end{itemize}

For this purpose, we consider the benchmark scenario described in Table~3 of \cite{Demartin:2014fia}, in which the 
coupling of the top-quark current to a ``Higgs Boson'' $X_0$ is described by the effective Lagrangian 
\begin{equation}
{\mathcal L}_0^t  = - \frac{m_t}{v} \, \overline{\psi} \left(\cos{\alpha} + i \sin{\alpha} \gamma_5 \right)\psi X_0 \, , \label{eq:effLag}
\end{equation}
where the angle $\alpha$ parameterizes the relative weight of the scalar and pseudoscalar couplings.
In Eq.~(\ref{eq:effLag}), $m_t$ indicates the mass of the the top quark and $v$ the vacuum expectation value. In the following we will identify $X_0$ with the SM Higgs boson $H$ when we consider a purely scalar coupling ($\alpha = 0$) while we will use the notation $X_0\to A$ when we consider a purely pseudoscalar coupling ($\alpha = \pi/2$). The part of the Lagrangian in Eq.~(\ref{eq:effLag}) which is proportional to $\gamma_5$ is odd under a CP transformation. While Eq.~(\ref{eq:effLag}) is not the most generic effective Lagrangian for this sector of the SM, it allows us to study in particular the effect of mixed scalar-pseudoscalr coupling on the shape of several differential distributions.

Experimentally, a full reconstruction of the momenta of the $t \bar{t} X_0$ final state is possible, and this allows one to measure  distributions which are differential with respect to the momenta of the final state massive particles. Four of these distributions play an important role in this work: the differential distribution with respect \emph{i)} to the invariant mass of the three massive particles in the final state, \emph{ii)} to the transverse momentum of the $X_0$ boson, \emph{iii)} to the invariant mass of the top pair, and \emph{iv)} to the transverse momentum of the top quark. The shapes of these distributions, with the exception of the transverse momentum of the top quark, show significant differences between the pure scalar and pseudoscalar cases at generator level. Consequently they can be used to experimentally probe the pseudoscalar component. Even though parton showering, detector acceptance, event selection and reconstruction are expected to considerably degrade these specific distributions and their discriminant power~\cite{Demartin:2014fia}, several studies strongly suggest that similar distributions involving full kinematic reconstruction can be measured~\cite{Santos:2015dja,AmorDosSantos:2017ayi}. In addition, these same studies indicate that new interesting angular distributions and asymmetries can be defined from the reconstructed top quarks and Higgs boson in order to discriminate the signal from the main irreducible backgrounds in dileptonic decays in an extremely efficient way, even after event reconstruction. Besides the fact that the full kinematic reconstruction of $t \bar{t} X_0$ events is extremely challenging, due to the presence of two undetected neutrinos in the dilepton channel, results show that angular distributions involving the reconstructed top quarks and Higgs boson and/or their decaying products are significantly different for the signal and the main irreducible backgrounds. A similar strategy can be implemented to probe the pseudoscalar component of the top quark Yukawa coupling, by exploring  new distributions, highly sensitive to the mixing angle. 

In order to obtain precise predictions for the associated production of a top pair and a $X_0$ boson  we write a parton level Monte Carlo which includes the resummation of soft emission corrections in the partonic threshold limit defined by $z = M^2/\hat{s} \to 1$, where $M$ is the invariant mass of the $t \bar{t} X_0$ final state and $\hat{s}$ is the partonic center of mass energy. The resummation of these corrections is carried to next-to-leading logarithmic (NLL) accuracy. Predictions to NLO+NLL accuracy for the total cross section and the four differential distributions listed above are obtained by matching the output of the parton level Monte Carlo with the NLO calculations obtained by using \mgamc.  

The paper is structured as follows. In Section~\ref{sec:numbers} we present predictions for the total cross section and differential distributions for $t \bar{t} X_0$ production considering the cases in which $X_0$ couples to the top quark as a scalar ($x_0 \to H$), pseudoscalar ($X_0 \to A$) and as a mixture   of the two. In Section~\ref{sec:expan} we discuss how the measurement of these quantities can be employed to extract information about the coupling of $X_0$ with the top. Section~\ref{sec:conclusions} contains our conclusions.

\section{Calculation and  Results \label{sec:numbers}}

The process of interest in this work is the associated production of a top pair and a boson $X_0$ which couples to the top quark according to the Lagrangian in Eq.~(\ref{eq:effLag}). From the calculational point of view, this process is equivalent to the associated production of a top pair and a Higgs boson in the SM. The resummation of the latter process in the partonic threshold limit was studied in detail in \cite{Broggio:2016lfj,Broggio:2015lya}. In particular, in \cite{Broggio:2016lfj} predictions for $t \bar{t} H$ production were obtained to NLO+NNLL accuracy. That calculation was carried out by means of an in-house parton level Monte Carlo code that was employed for the numerical evaluation of the resummation formula to NNLL accuracy. The NNLL calculations were then matched to NLO results obtained by using \mgamc. Our goal in this paper is to calculate to NLL accuracy the $t \bar{t} X_0$ production process of an arbitrary mixture of scalar and pseudoscalar coupling of $X_0$. We then match the results obtained to NLO calculations carried out by using the the model package for \mgamc~developed for the work published in \cite{Demartin:2014fia}.

The resummation formula which we employ is derived along the lines of the one discussed in \cite{Broggio:2016lfj} for the SM case  and of the ones obtained in \cite{Broggio:2016zgg,Broggio:2017kzi} for $t \bar{t} W$ and $t \bar{t} Z$ production, respectively.  For this reason, we do not repeat a detailed description of the resummation formula and of the resummation technique here\footnote{For an introduction to the soft-collinear effective theory methods employed in this approach to partonic threshold resummation see for example \cite{Becher:2014oda}.}. We simply point out that in order to carry out the resummation to NLL accuracy one needs to evaluate the anomalous dimensions, soft functions and hard functions to LO only. The anomalous dimensions and the soft functions are identical to the ones employed in the SM calculation. The LO hard function for the quark-annihilation and gluon fusion  partonic production channels can be readily obtained with a simple analytic calculation. The numerical evaluation of the resummed cross section is carried out in Mellin moment space, as it was done in \cite{Broggio:2017kzi,Broggio:2016zgg,Broggio:2016lfj}. Physical results are then obtained with a numerical inverse Mellin transform carried out by employing the \emph{Minimal Prescription} to avoid problems related to the presence of the Landau pole \cite{Catani:1996yz}.

When combining resummed calculations to fixed order calculations, it is important to avoid the double counting of terms included in both approaches. This is achieved through the procedure of ``matching''. In the case considered in this work, the matching procedure can be summarized in a schematic equation, which we write down here for the case of the total cross section:
\begin{align}
\sigma^{{\footnotesize \text{NLO+NLL}}} = \sigma^{{\footnotesize \text{NLO}}} + \left[\sigma^{{\footnotesize \text{NLL}}} -  \sigma^{{\footnotesize \text{NLL expanded to NLO}}}\right]
\label{eq:matching}
\end{align}
In Eq.~(\ref{eq:matching}) the terms within square brackets give contributions starting at NNLO.

Resummed calculations to a given logarithmic accuracy have a residual dependence on three non-physical scales: the factorization scale $\mu_f$, which appears also in fixed order calculations, the soft scale $\mu_s$ which characterizes the soft gluon emission, and the hard scale $\mu_h$ which is the scale that characterizes the hard scattering. The hard and soft scales appear only in the resummed partonic cross section. The soft function evaluated at the scale $\mu_s$ and the hard function evaluated at the scale $\mu_h$ are free from large logarithms. By solving the Renormalization Group Equations satisfied by the soft and hard functions the partonic cross section is evolved to the factorization scale $\mu_f$ and then convoluted with the partonic luminosity. 
In fixed order calculations, the residual perturbative uncertainty associated to corrections at perturbative orders which are higher than the one considered in the calculation is estimated by varying the default choice for the factorization scale by a factor of two. To be specific, if one indicates the default choice for the factorization scale with $\mu_{f,0}$, the scale is then varied in the interval $\mu_f \in [\mu_{f,0}/2,2 \mu_{f,0}]$; subsequently, the scale uncertainty affecting the fixed order result is taken to be the interval included between the largest and smallest values of the observable obtained while varying the factorization scale.  In resummed calculations, also the hard and soft scales are varied by a factor of two around their default value, indicated by $\mu_{s,0}$ and $\mu_{h,0}$ respectively: $\mu_s \in [\mu_{s,0}/2,2 \mu_{s,0}]$ and  $\mu_h \in [\mu_{h,0}/2,2 \mu_{h,0}]$. 
The perturbative scale uncertainty affecting resummed calculations is estimated by varying separately 
the hard, soft, and factorization scales and by subsequently combining the uncertainties in quadrature. 
In particular, for any observable $O$ (which can be the total cross section or the value of the differential cross section in a particular bin), one evaluates the quantities
\begin{align}
\Delta O^+_i &= \max\{O\left(\kappa_i=1/2\right),O\left(\kappa_i=1\right), O\left(\kappa_i=2\right)\} \nonumber \\
& -
O\left(\kappa_i=1\right) \, , \nonumber \\
\Delta O^-_i &= \min\{O\left(\kappa_i=1/2\right),O\left(\kappa_i=1\right), O\left(\kappa_i=2\right)\} \nonumber \\
& -
O\left(\kappa_i=1\right) \,. \label{eq:Deltas}
\end{align}
In Eqs.~(\ref{eq:Deltas}) we introduced the quantities $\kappa_i = \mu_{i}/\mu_{i,0}$ where the index $i \in \{s,h,f\}$. The three quantities $\Delta O^+_i$ ($\Delta O^-_i$) are then combined in quadrature in order to obtain the scale uncertainty above (below) the central value.

The dependence of the physical predictions on the factorization, hard and soft scales is expected to become progressively smaller as more perturbative orders are added to fixed order calculations and a higher logarithmic accuracy is reached in resummed calculations. Since we obtain predictions at NLO+NLL accuracy, it is important to choose carefully the default values for the hard, soft and factorization scales. For all of the scales we choose dynamic default values depending on $t \bar{t} X_0$ invariant mass. With the method chosen for the resummation of the soft emission corrections the choice of the hard and soft scales is straightforward and is dictated by the kind of scale dependent logarithms found in the hard and soft function respectively. This leads to the choice $\mu_{h,0} = M$ and $\mu_{s,0} = M /\bar{N}$, where $\bar{N} = N e^{\gamma_\text{E}}$ and $N$ is the Mellin moment parameter while $\gamma_\text{E}$ indicates the Euler-Mascheroni constant. For the SM case, the choice of a suitable central value for the factorization scale was discussed in \cite{Broggio:2016lfj}. The total $t \bar{t} H$ total cross section calculated as a function of $\mu_f$ at NLO, NLO+NLL and NLO+NNLL shows that these three curves intersect each other at $\mu_f/ M \sim 0.5$ (see Fig.~1 in \cite{Broggio:2016lfj}) while the curves differ significantly for much smaller or much larger values of $\mu_f$. This fact motivated the choice $\mu_{f,0} = M/2$ in \cite{Broggio:2016lfj}. The process considered in this work differs from the SM $t \bar{t} H$ production simply because of the presence of the pseudoscalar coupling in Eq.~(\ref{eq:effLag}). Since the pseudoscalar coupling is not expected to change the scale behavior of the cross section, we set  $\mu_{f,0} = M/2$ also in this work.

\begin{table}[t]
	\begin{center}
		\def\arraystretch{1.3}
		\begin{tabular}{|c|c||c|c|}
			\hline $M_W$ & $80.419$~GeV & $m_t$ & $173$~GeV\\ 
			\hline $M_Z$ &  $91.188$~GeV & $m_H$ & $125$~GeV \\ 
			\hline $1/\alpha$ & $137.036$ & $\alpha_s \left(M_Z\right)$ & from MMHT 2014 PDFs \\ 
			\hline 
		\end{tabular} 
		\caption{Input parameters employed throughout the calculation. \label{tab:tabGmu}}
	\end{center}
\end{table}

Finally, the calculations of the total cross section and differential distributions presented in the rest of this paper were obtained by employing the input parameters listed in Table~\ref{tab:tabGmu}. The input parameters chosen are the same ones employed in~\cite{AmorDosSantos:2017ayi}. Throughout the paper we employ MMHT 2014 PDFs \cite{Harland-Lang:2014zoa}; in particular, we employ LO PDFs in LO calculations and NLO PDFs in NLO and NLO+NLL calculations.

\subsection{Total Cross Section}

\begin{table}[t]
	\begin{center}
		\def\arraystretch{1.5}
		\begin{tabular}{|c|c|c|c|}
			\hline  order & PDF order & $\alpha$ & $\sigma$ [fb]\\ 
			\hline LO & LO & $0$ & $378.7^{+120.6 \, (32 \%)}_{-85.2 \, (22 \%)} $ \\
			\hline LO & LO & $\pi/2$ & $142.4^{+50.1 \, (35 \%) }_{-34.6 \, ( 24 \%)} $ \\
			\hline 
		         \hline NLO & NLO & $0$ & $ 475.0^{+47.3 \, (10 \%)}_{-51.9 \, (11 \%)} $ \\
		         \hline NLO & NLO & $\pi/2$ & $ 192.4^{+23.3 \, (12 \%)}_{-24.3\, (13 \%)} $ \\
			\hline \hline NLO+NLL & NLO& $0$  & $480.3^{+57.8\, (12 \%)}_{- 15.6 (3.2 \%)}$ \\
			\hline NLO+NLL & NLO& $\pi/2$  & $199.6^{+17.7 (8.9 \%)}_{-8.4 (4.2 \%)}$ \\
								\hline 
		\end{tabular} 
		\caption{Total cross section for $t \bar{t} X_0$ production at the
                  LHC with $\sqrt{s} = 13$~TeV and MMHT 2014 PDFs. The default value
                    of the factorization scale is $\mu_{f,0}=M/2$, and
                    the uncertainties are estimated through 
                    variations of this scale (and of the hard and soft  scales
                    $\mu_s$ and $\mu_h$ in resummed calculations). 
\label{tab:CSHp13hM}}
	\end{center}
\end{table}

We start by discussing results for the total cross section. Table~\ref{tab:CSHp13hM} lists values for the total cross section for the purely scalar coupling case ($\alpha = 0$) and for the purely pseudoscalar case ($\alpha = \pi/2$). The table reports numbers for LO, NLO and NLO+NLL calculations. The residual perturbative uncertainty is estimated by varying the factorization scale (in all calculations) and the hard and soft scales (in NLO+NLL calculations) as explained in the previous section. 
Table~\ref{tab:CSHp13hM} shows that the NLO corrections increase significantly the central value of the total cross section for the two values of $\alpha$ considered. Furthermore, as expected, the NLO scale uncertainty is significantly smaller than the LO one. The resummation of soft emission corrections to NLL accuracy results into a small increase of the central value of the cross section, of the order of a few percent. The scale uncertainty interval obtained by varying the three scales in the resummed cross section is smaller than the scale uncertainty interval affecting the NLO calculation for both values of $\alpha$. We notice that the resummation effect is larger for the purely pseudoscalar case than for the purely scalar one. This might be due to the fact that in the purely pseudoscalar case the large invariant mass region contributes, in proportion, to a larger fraction of the total cross section with respect to the purely scalar case. This fact can be noticed by looking at the invariant mass distribution shown in the upper-left plot of Figure~\ref{fig:NLOdist}. In the invariant mass tail the soft emission corrections dominate the cross section, hence resummation has larger effects in that region.
The PDF uncertainty on the NLO results was evaluated separately by means of \mgamc~by considering the \verb|MMHT2014nlo68cl| PDF set. It is of the order or $\sim \pm 3 \%$. The PDF uncertanty affecting the NLO+NLL result is expected to be of the same magnitude.

%
\begin{table}[t]
	\begin{center}
		\def\arraystretch{1.5}
		\begin{tabular}{|c|c|c|c|}
			\hline  order & PDF order & $\alpha$ & $\sigma$ [fb]\\ 
		         \hline NLO & NLO & $0$ & $ 475.0^{+47.3 \, (10 \%)}_{-51.9 \, (11 \%)} $ \\
\hline NLO & NLO & $\pi/6$ & $ 404.4^{+41.3 \, (10 \%)}_{-45.0 \, (11 \%)} $ \\
\hline NLO & NLO & $\pi/4$ & $ 333.7^{+ 35.3\, (11 \%)}_{- 38.1\, (11 \%)} $ \\
\hline NLO & NLO & $\pi/3$ & $ 263.1^{+ 29.3\, (11 \%)}_{- 31.2	\, (12 \%)} $ \\
\hline NLO & NLO & $\pi/2$ & $ 192.4^{+23.3 \, (12 \%)}_{-24.3\, (13 \%)} $ \\
\hline \hline NLO+NLL & NLO& $0$  & $480.3^{+57.8\, (12 \%)}_{- 15.6 (3.2 \%)}$ \\
\hline NLO+NLL & NLO& $\pi/6$  & $410.1^{+47.4\, (12 \%)}_{- 12.4  (3.0 \%)}$ \\
\hline NLO+NLL & NLO& $\pi/4$  & $339.9^{+37.0\, (11 \%)}_{- 9.8  (2.9 \%)}$ \\
\hline NLO+NLL & NLO& $\pi/3$  & $269.8^{+ 27.0 \, (10 \%)}_{- 8.4  (3.1 \%)}$ \\
\hline NLO+NLL & NLO& $\pi/2$  & $199.6^{+17.7 (8.9 \%)}_{-8.4  (4.2 \%)}$ \\
								\hline 
		\end{tabular} 
		\caption{Total cross section for $t \bar{t} X_0$ production at the
                  LHC with $\sqrt{s} = 13$~TeV as a function of the angle $\alpha$. 
\label{tab:CSHp13hMalpha}}
	\end{center}
\end{table}

\begin{figure*}
\begin{center}
\includegraphics[width=12.2cm]{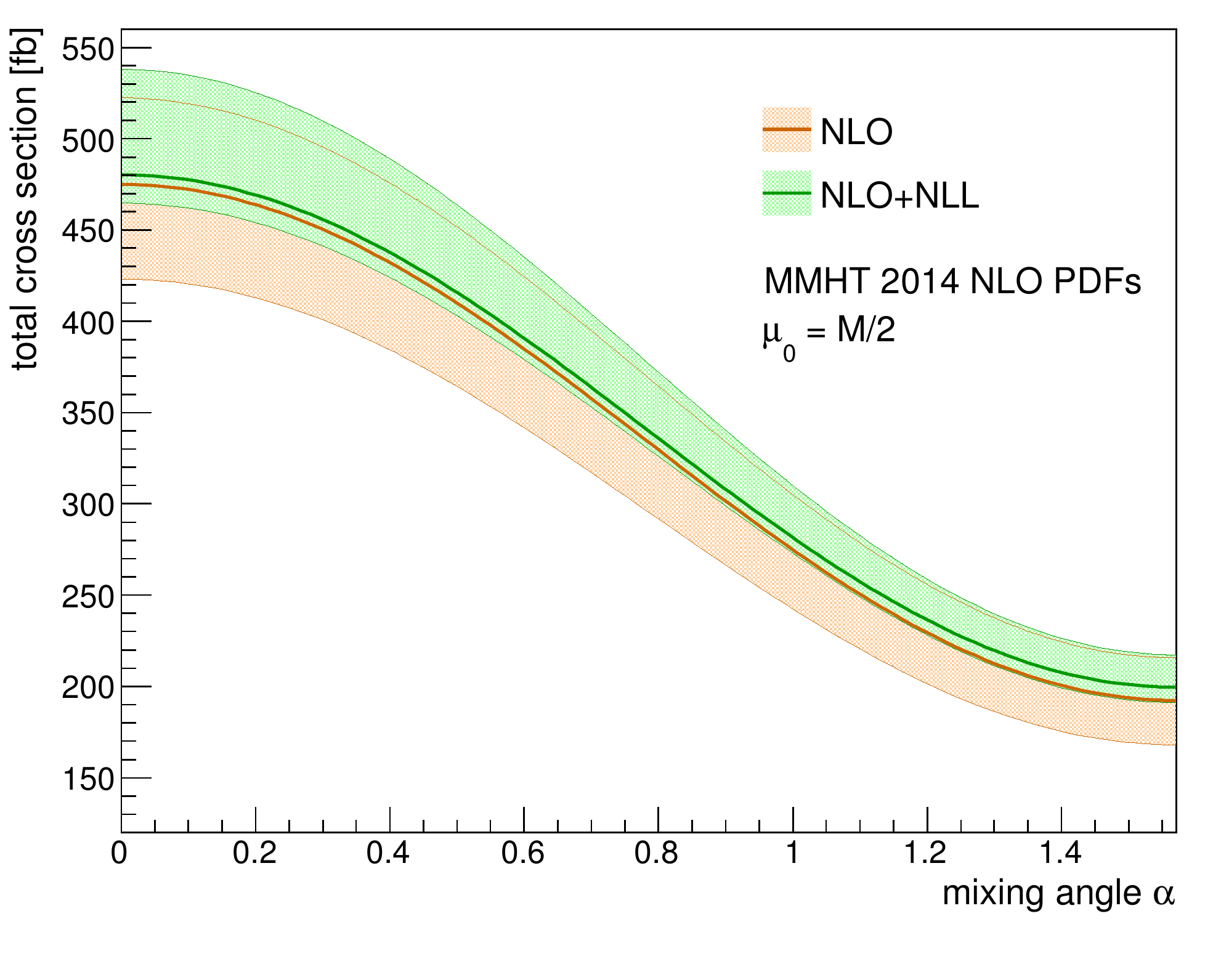}
\caption{Dependence of the total cross-section on the mixing angle $\alpha$.}
\label{fig:TOTCSdependenceal}
\end{center}
\end{figure*}

The result for other values of the mixing angle $\alpha$ can be obtained by employing the relation
\begin{equation}
\sigma(\alpha) = \sigma_H \cos^2 \alpha + \sigma_A \sin^2 \alpha \, .
\label{eq:sigma}
\end{equation}
In Eq.~(\ref{eq:sigma}), $\sigma_H$ indicates the total cross section for a pure SM like coupling, while $\sigma_A$ indicates the cross section for a pure pseudoscalar coupling. 
At NLO, the scale uncertainty at an arbitrary angle can be obtained in a similar way by exploiting the fact that the upper edge of the scale uncertainty band corresponds to the NLO calculation in which the factorization scale is set equal to $\mu_{f, 0}/2$, while the lower edge corresponds to the calculation in which one sets $\mu_f = 2 \mu_{f, 0}$. 
Eq.~(\ref{eq:sigma}) can also be employed to obtain the central value of the NLO+NLL cross section at an arbitrary angle. However, since the estimate of the scale uncertainty at NLO+NLL accuracy involves the independent variation of three scales, it cannot be obtained directly from the results of Table~\ref{tab:CSHp13hM} and it must be evaluted separately. Table~\ref{tab:CSHp13hMalpha} lists the values of the NLO and NLO+NLL cross section and of their scale uncertianty for a few values of the angle $\alpha$. Figure~\ref{fig:TOTCSdependenceal} shows the dependence of the total cross-section on the mixing angle. For all values of $\alpha$ the NLO+NLL cross section is a few femtobarn higher than the NLO one, while the scale uncertainty intervals at NLO+NLL are slightly smaller than the corresponding NLO intervals.

\subsection{Differential Distributions}

In this section we consider four differential distributions for the $t \bar{t} X_0$ production process:
\begin{itemize}
\item[\emph{i)}] Distribution differential with respect to the $t \bar{t} X_0$ invariant mass $M$,
\item[\emph{ii)}] distribution differential with respect to the transverse momentum of the $X_0$ scalar $p_T^X$,
\item[\emph{iii)}] distribution differential with respect to the top-pair invariant mass $M_{t \bar{t}}$, and, finally,
\item[\emph{iv)}] distribution differential with respect to the transverse momentum of the top quark, $p_T^t$.
\end{itemize}
Our goal is to see
if these distributions, and in particular their shapes, are sensitive to the scalar and/or pseudoscalar nature of the $X_0$ coupling. As for the total cross section, we consider here the case $\alpha = 0$ (SM-like Higgs boson) and $\alpha = \pi/2$ (pure pseudoscalar coupling). The value of the differential distributions for arbitrary $\alpha$ can be obtained by combining the distribution at  $\alpha = 0$ and $\alpha = \pi/2$ as shown in Eq.~(\ref{eq:sigma}) for the case of the total cross section.
In addition, we want to check if the shape of these distributions is stable with respect to the inclusion of soft gluon emission corrections through the resummation process. 

\begin{figure*}
	\begin{center}
		\begin{tabular}{cc}
			\includegraphics[width=7.0cm]{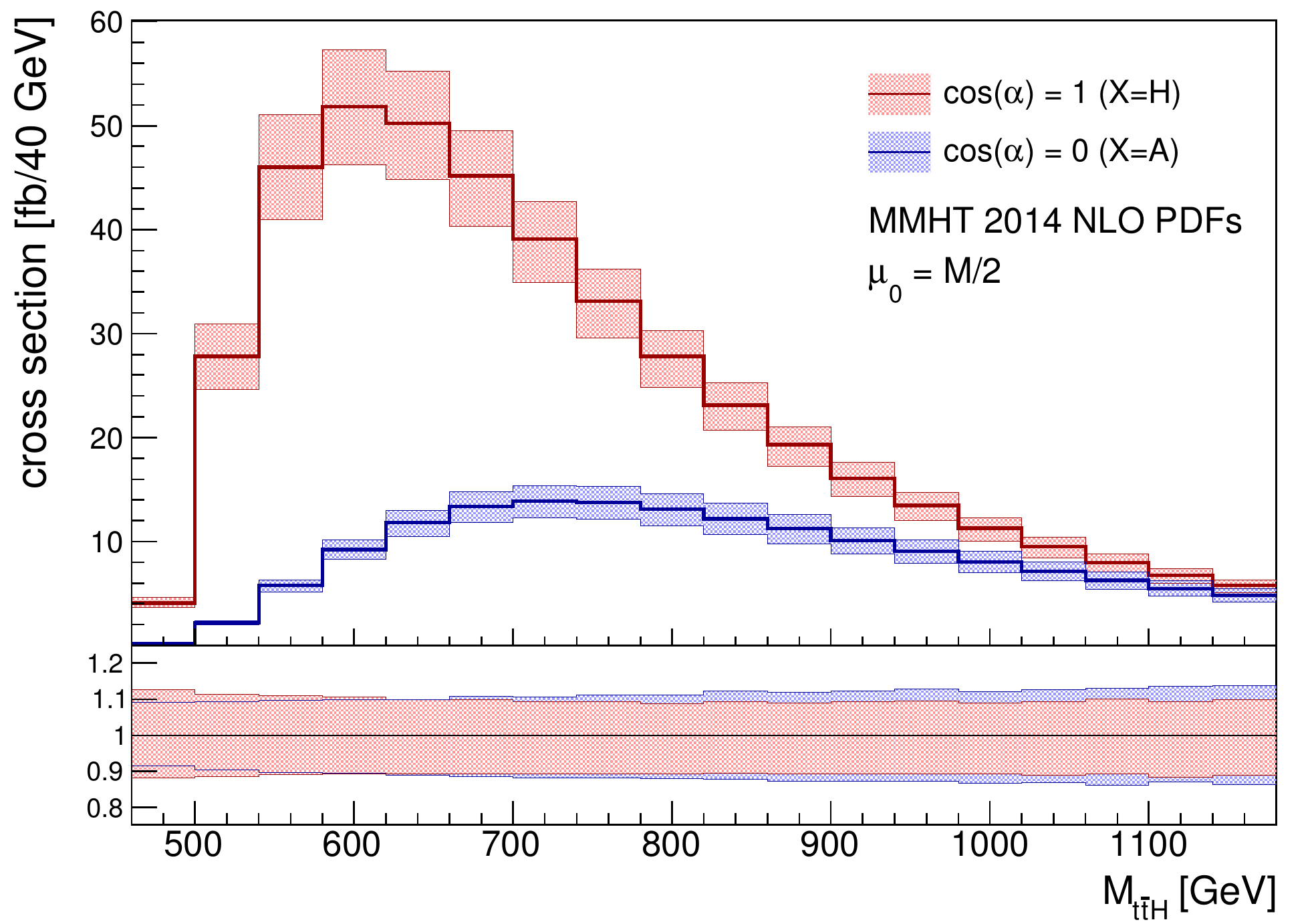} & \includegraphics[width=7.0cm]{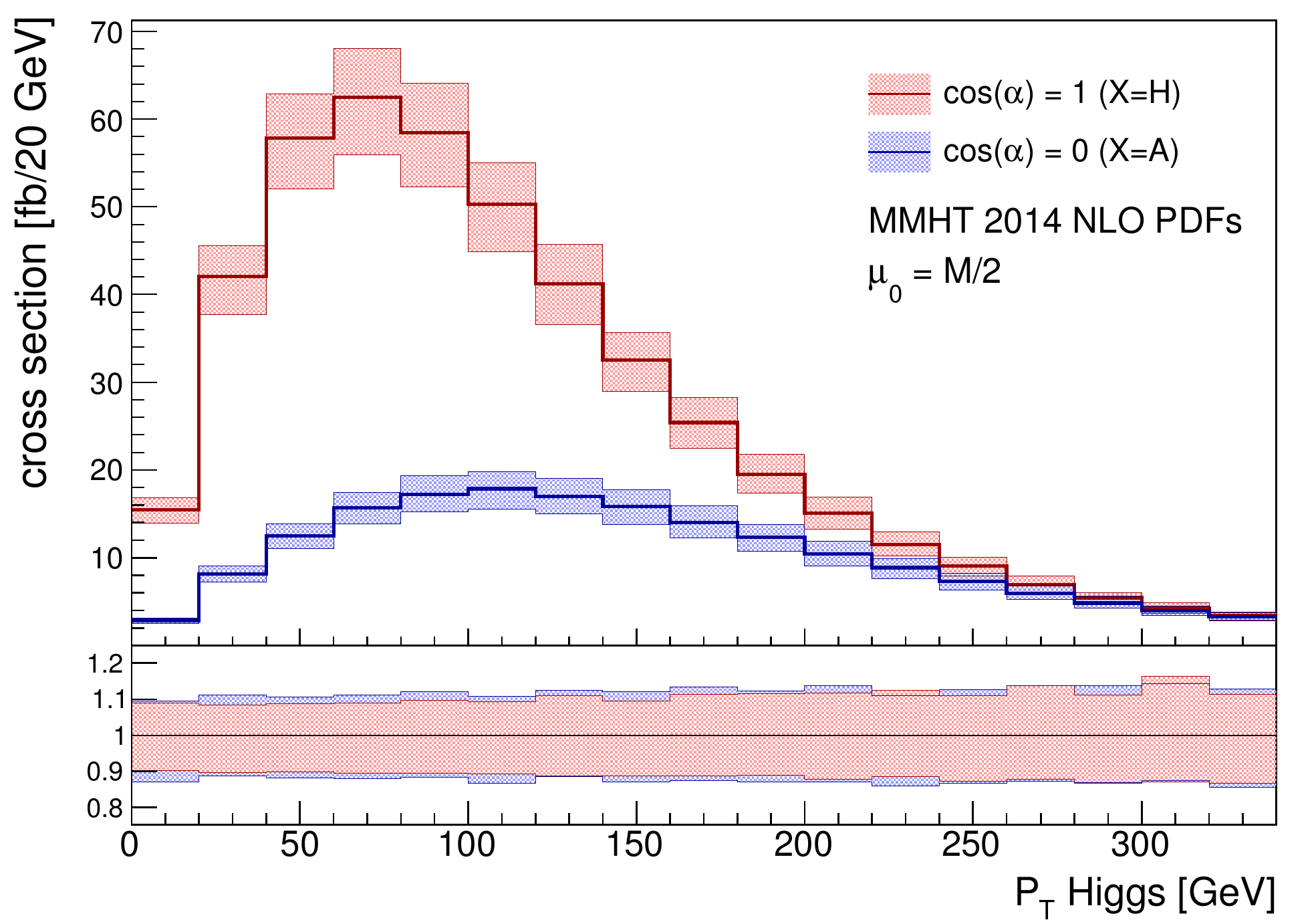} \\
			\includegraphics[width=7.0cm]{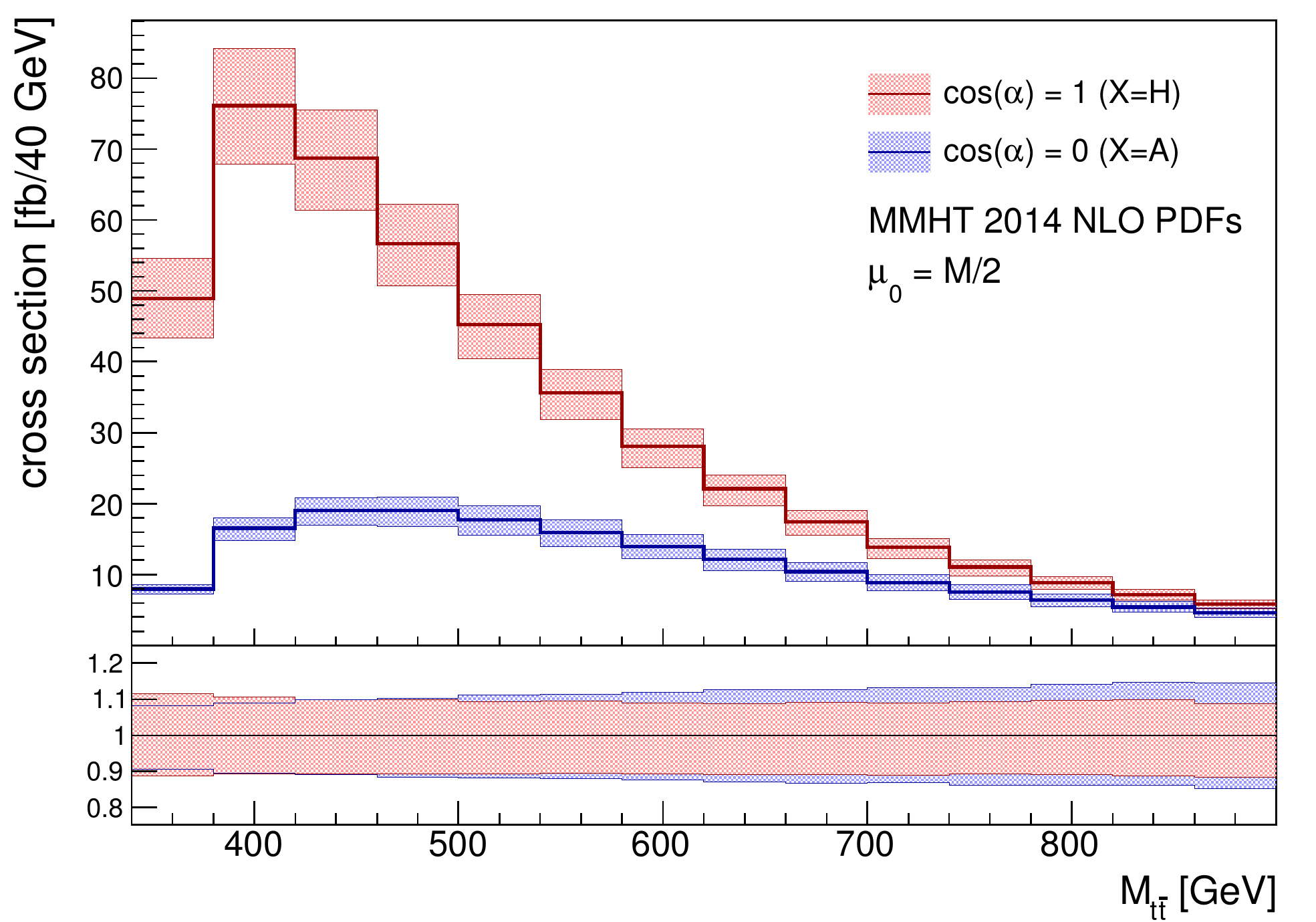} & \includegraphics[width=7.0cm]{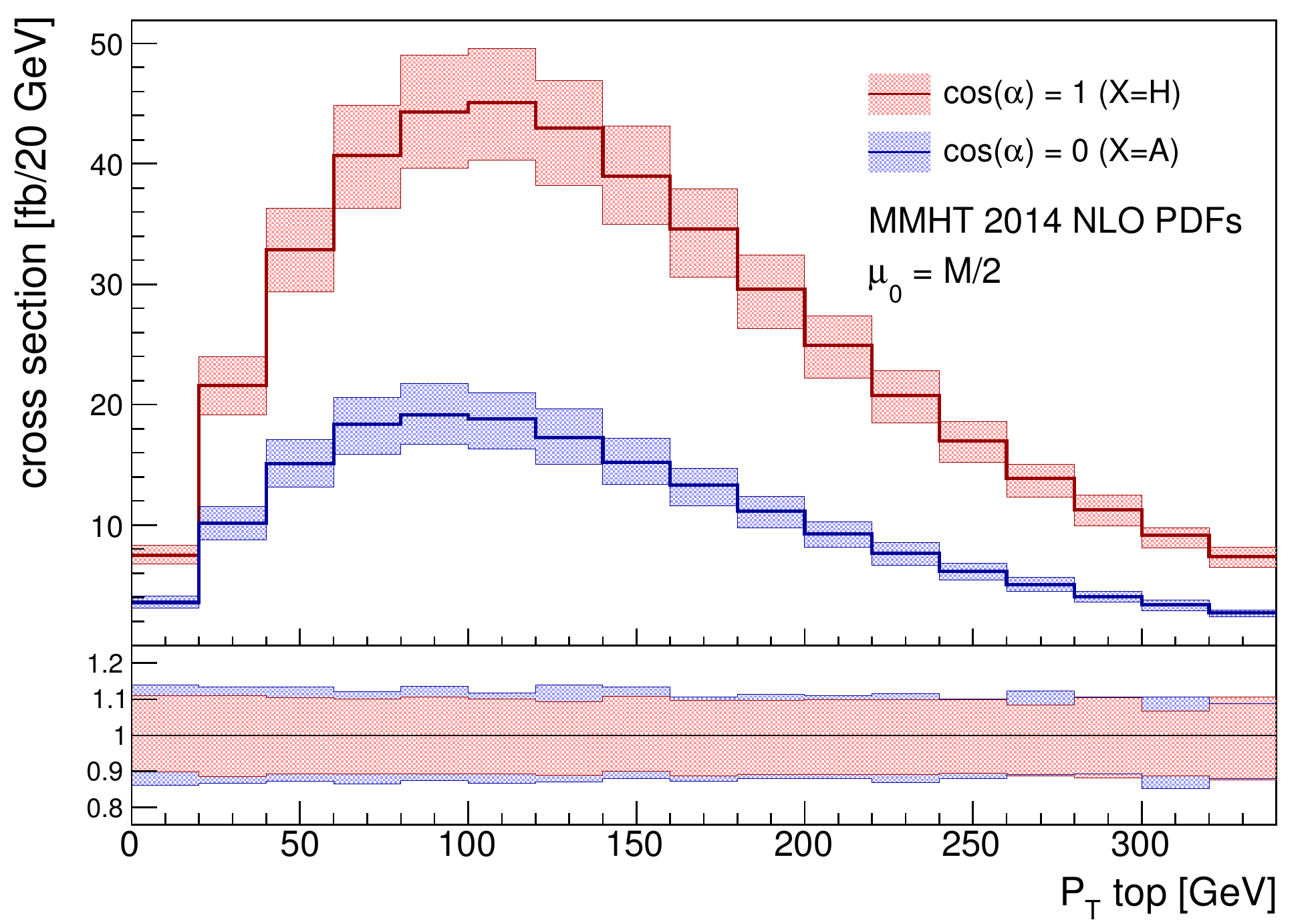} \\
		\end{tabular}
	\end{center}
    \vspace*{-4mm}
	\caption{Differential distributions evaluated at NLO. The default
		factorization scale is chosen as $\mu_{f,0}=M/2$, and the
		uncertainty bands are generated through scale variations as
		explained in the text. \label{fig:NLOdist}} 
\end{figure*}

These differential distributions evaluated to NLO with \mgamc~are shown in Figure~\ref{fig:NLOdist}. The bands represent the residual perturbative uncertainty evaluated through the variation of the factorization scale as explained above. At this stage, we made no attempt to account for the PDFs uncertainty. The bottom part of each panel shows the width of each bin divided by its central value, and allows one to assess the relative scale uncertainty of the bands for the scalar and pseudoscalar cases. We see that the relative scale uncertainties for the cases $\alpha = 0$ and $\alpha = \pi/2$ are very similar. 

\begin{figure*}
	\begin{center}
		\begin{tabular}{cc}
			\includegraphics[width=7.0cm]{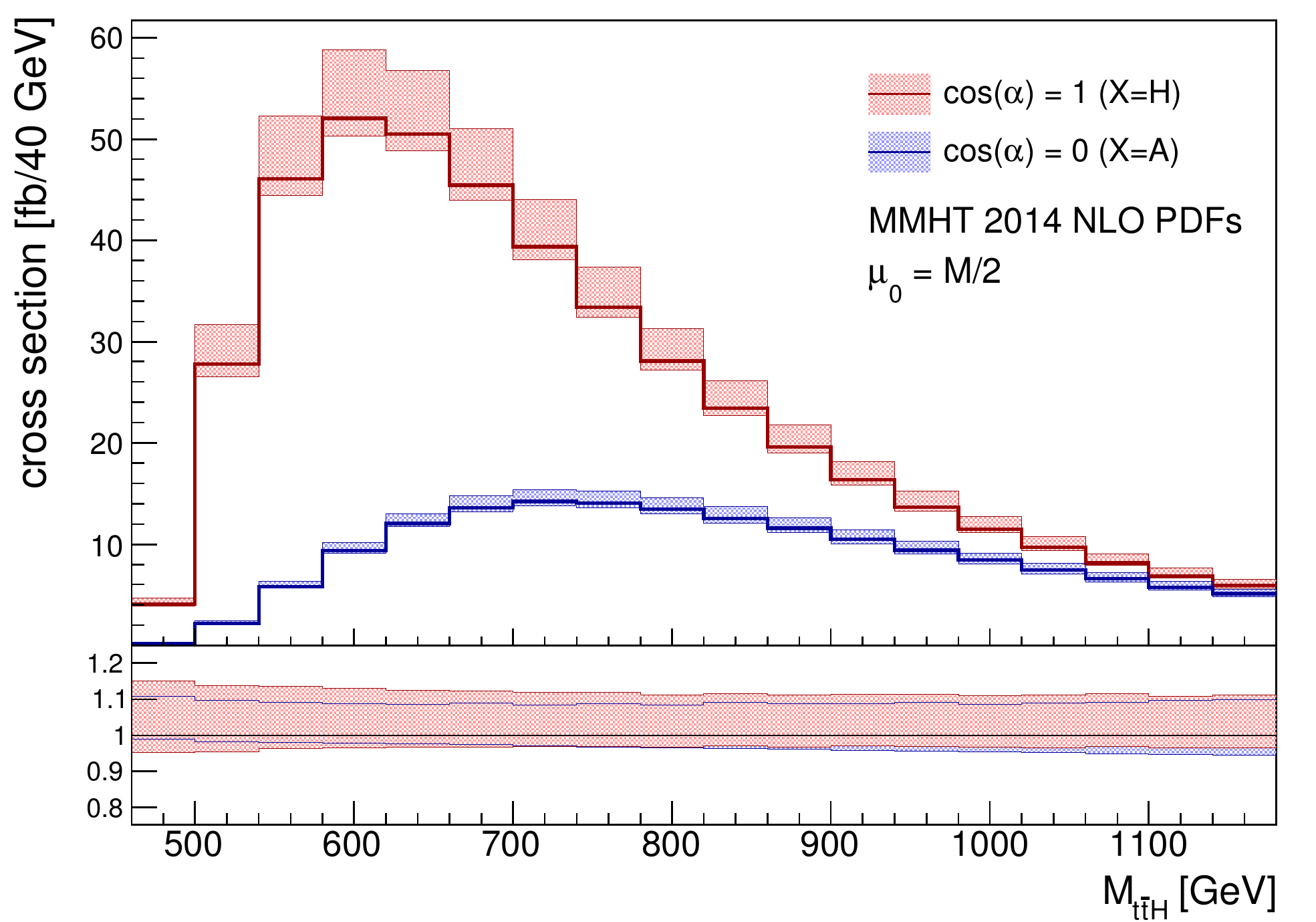} & \includegraphics[width=7.0cm]{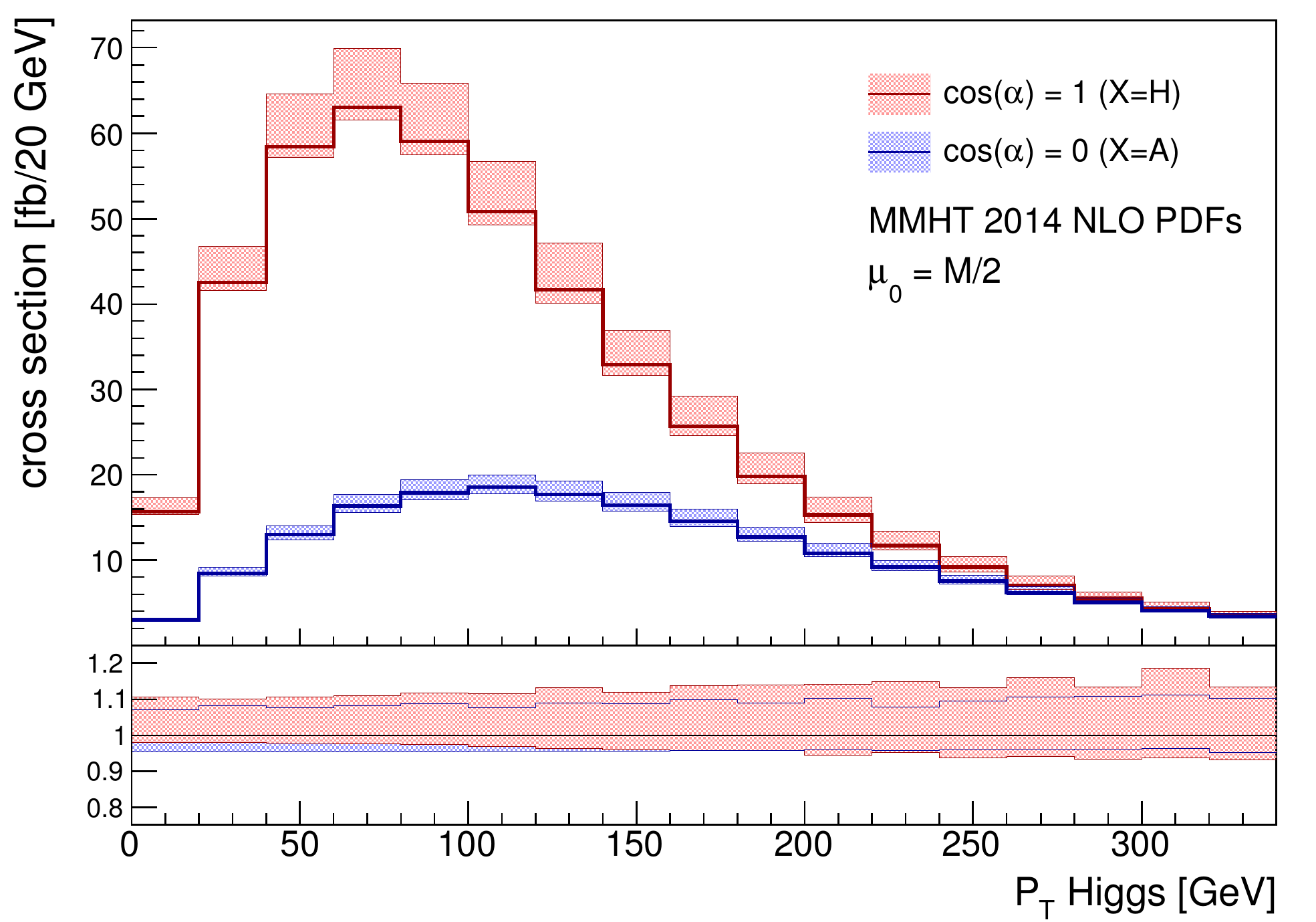} \\
			\includegraphics[width=7.0cm]{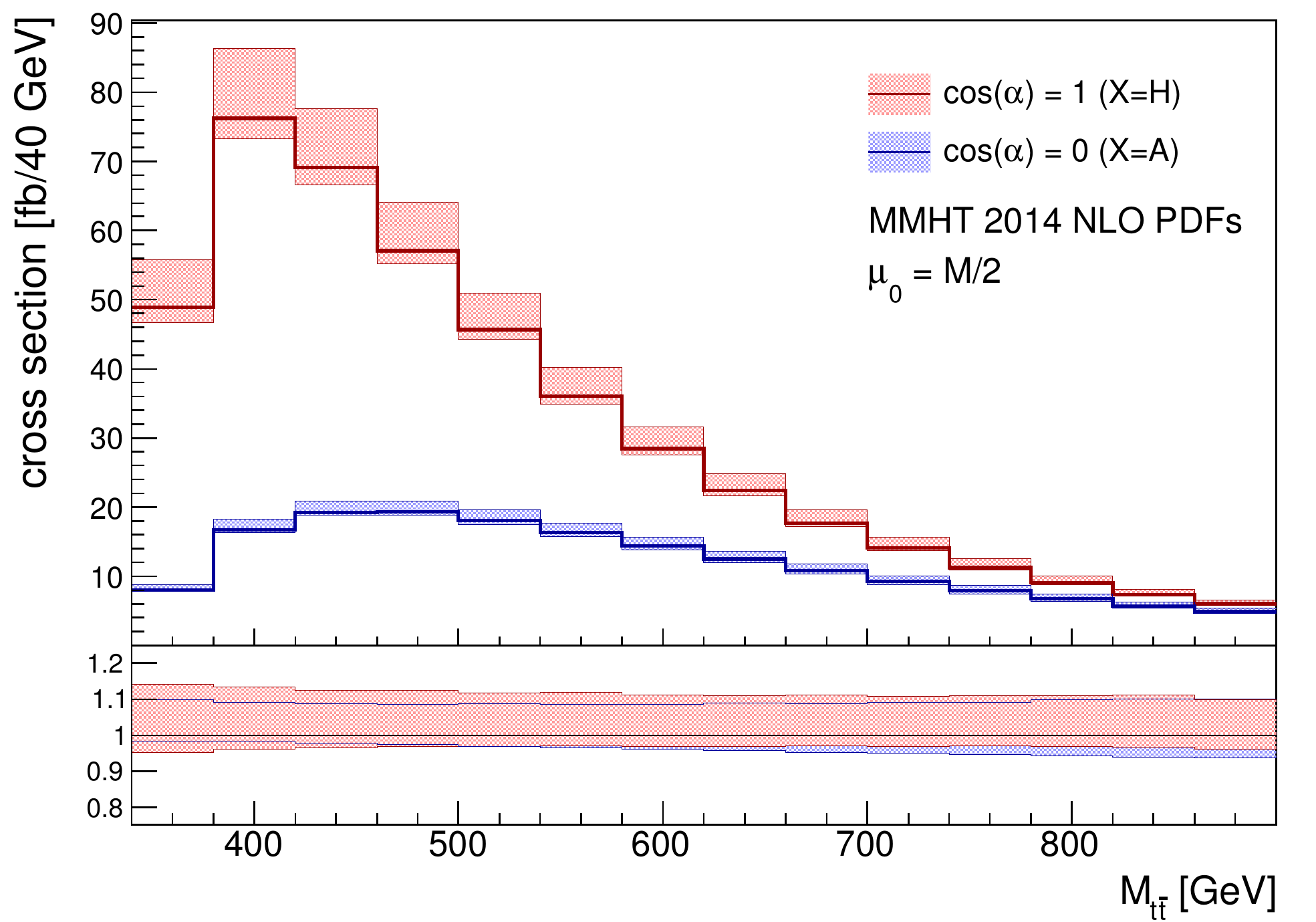} & \includegraphics[width=7.0cm]{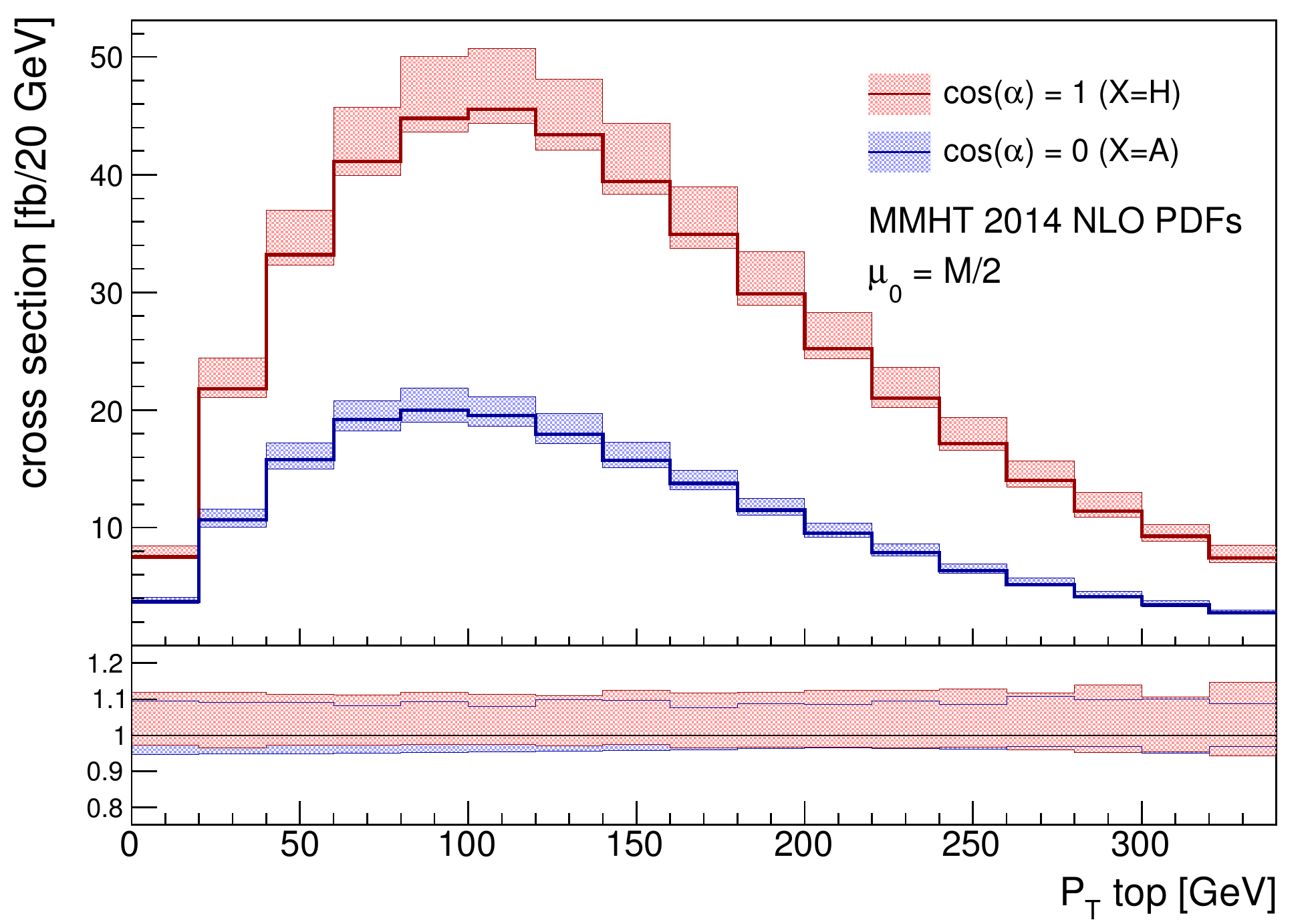}
\\
		\end{tabular}
	\end{center}
    \vspace*{-4mm}
	\caption{Differential distributions evaluated at NLO+NLL. The
		uncertainty bands are generated through scale variations as
		explained in the text. \label{fig:NLONLLdist}} 
\end{figure*}

\begin{figure*}
	\begin{center}
		\begin{tabular}{cc}
			\includegraphics[width=7.0cm]{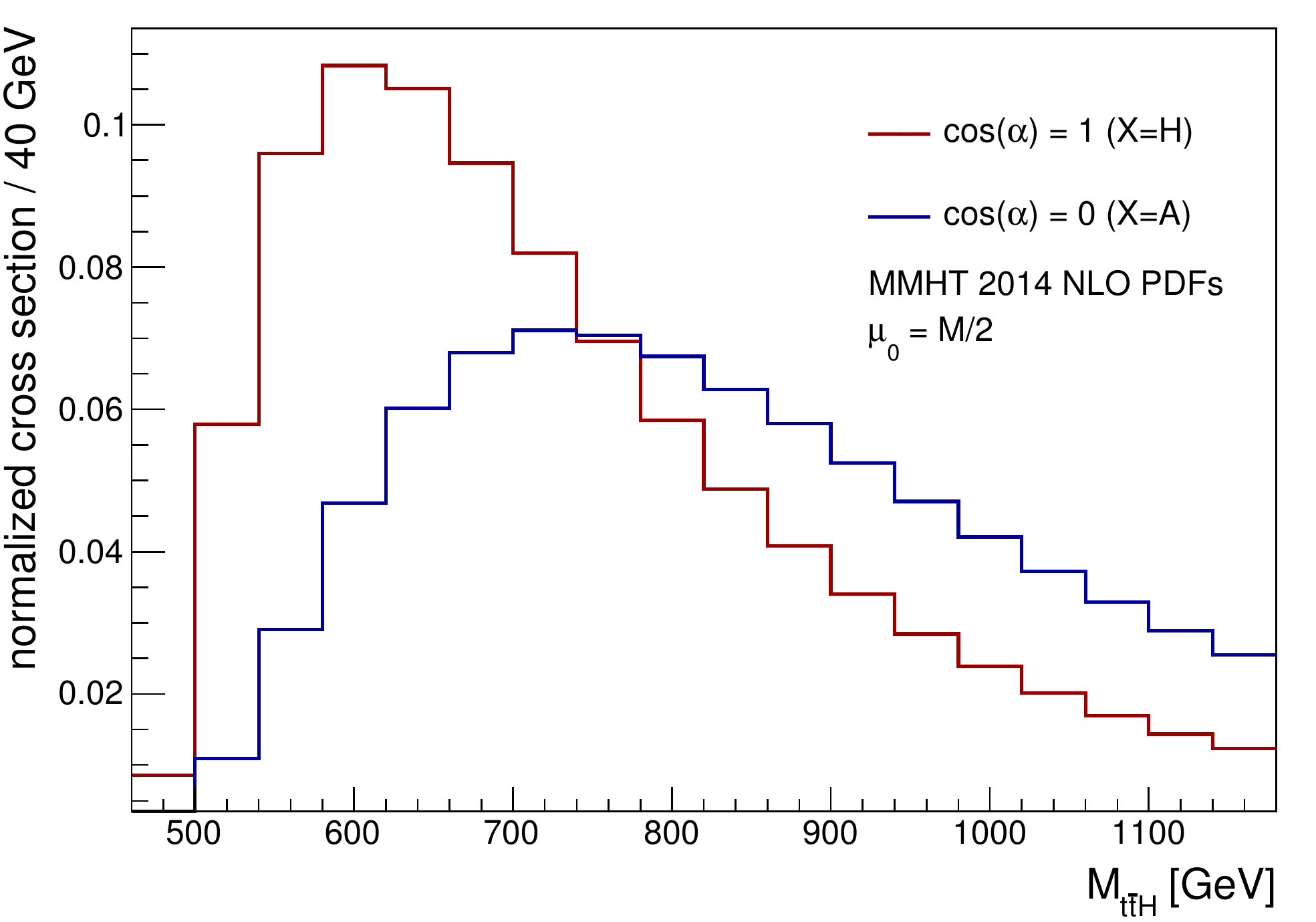} & \includegraphics[width=7.0cm]{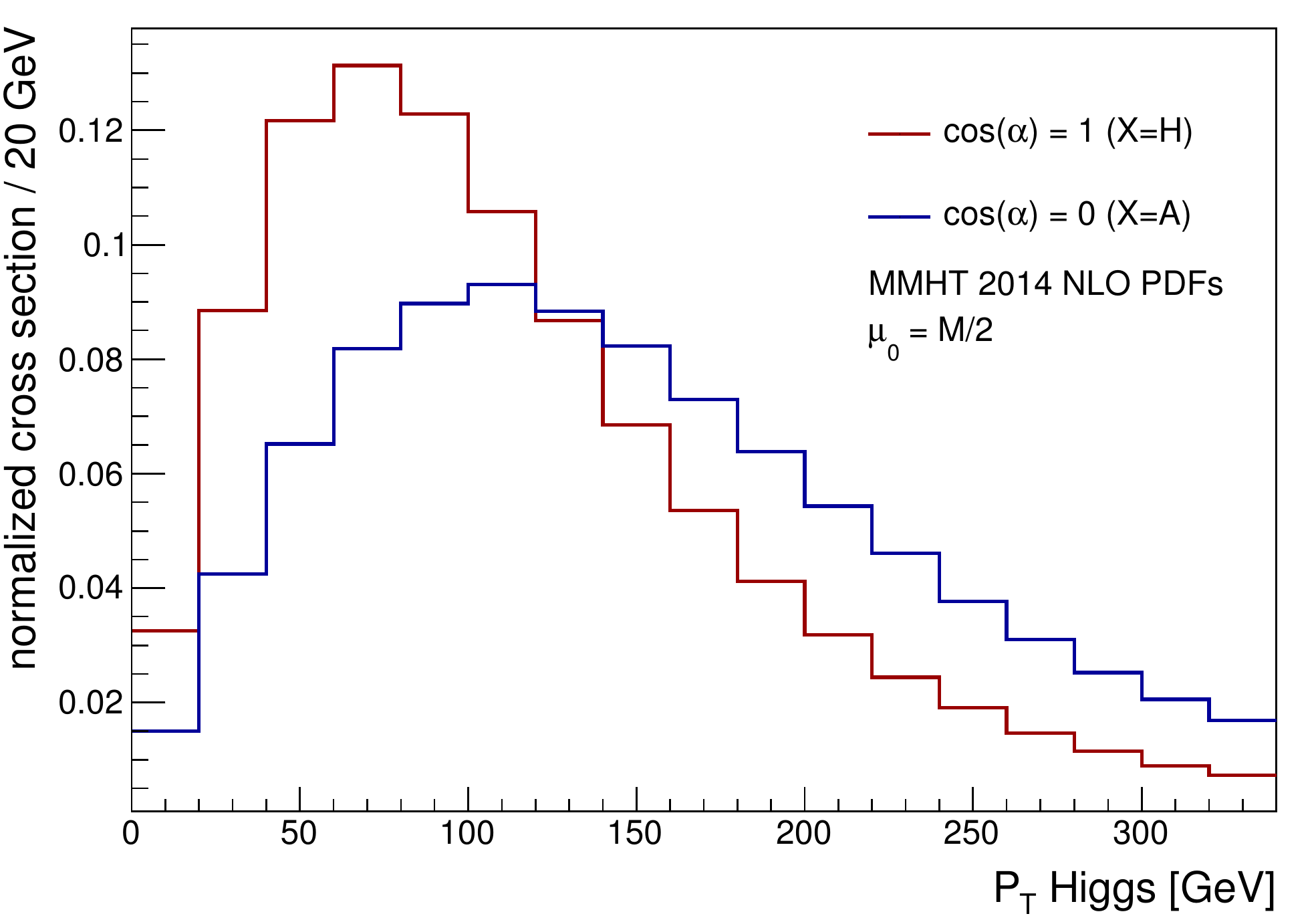} \\
			\includegraphics[width=7.0cm]{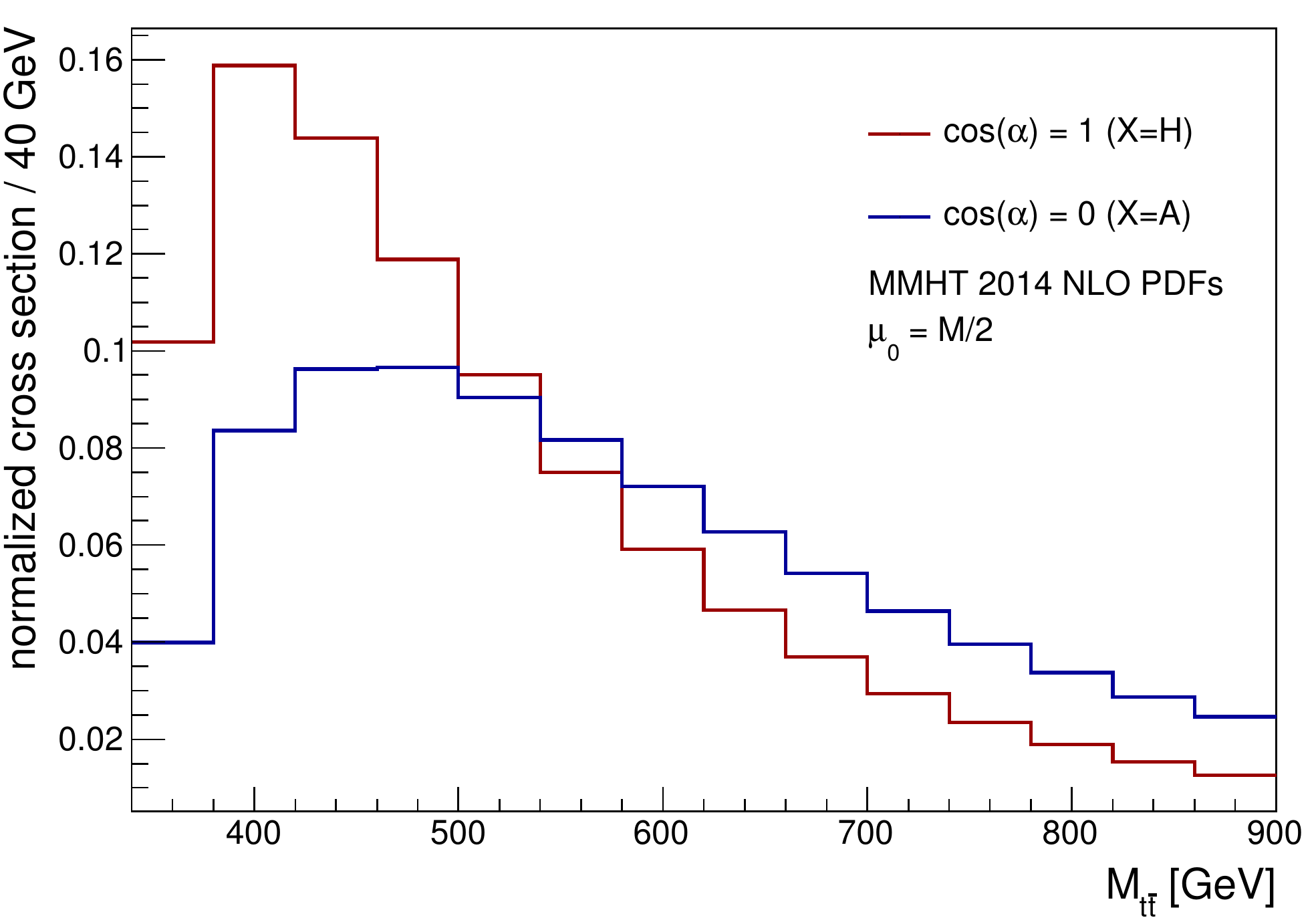} & \includegraphics[width=7.0cm]{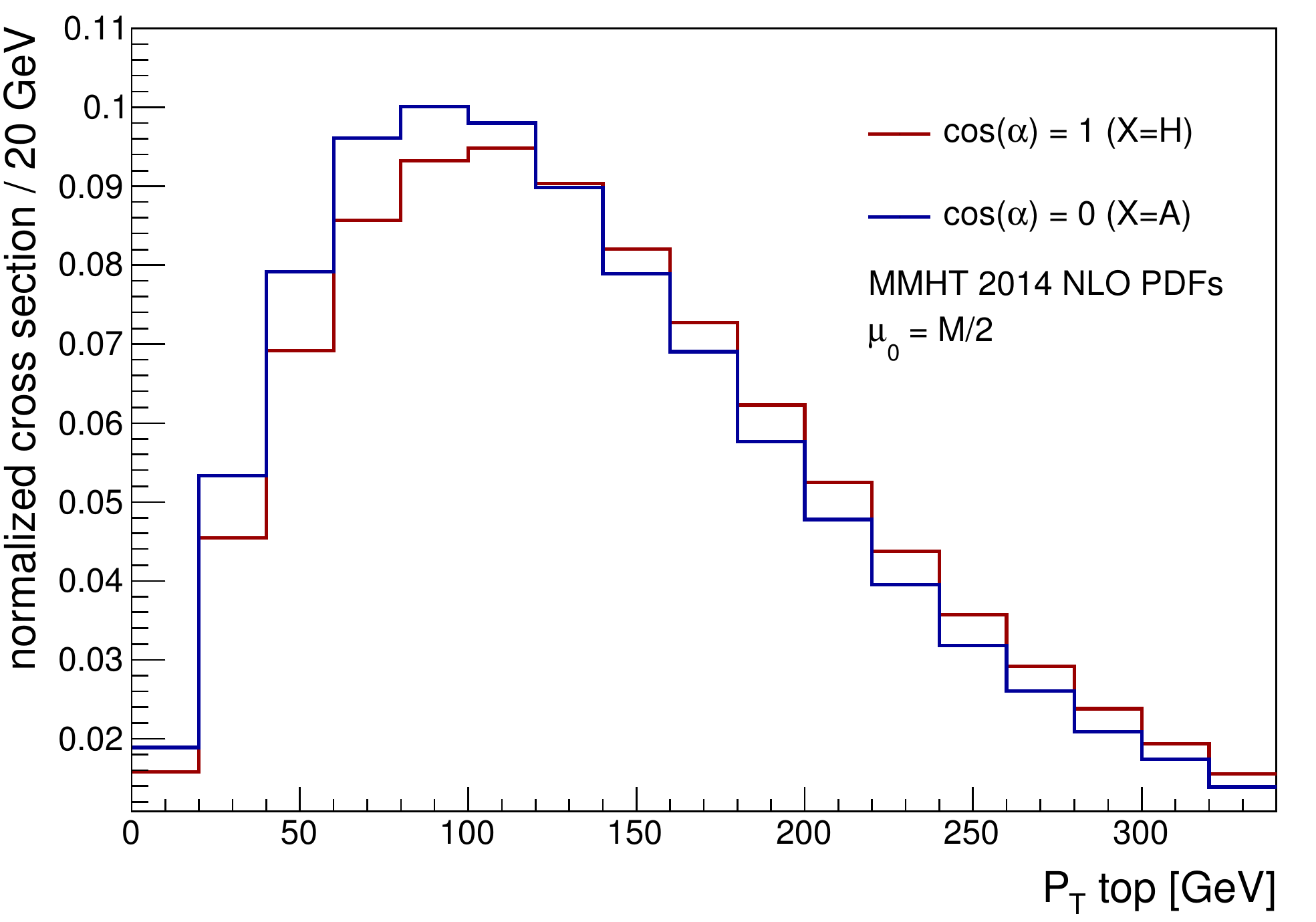}
\\
		\end{tabular}
	\end{center}
    \vspace*{-4mm}
	\caption{Normalized differential distributions evaluated at NLO+NLL. The factorization scale is set to $\mu_f = M/2$. \label{fig:NLONLLnorm}} 
\end{figure*}

The distributions presented in Figure~\ref{fig:NLONLLdist} are the same distributions shown in Figures~\ref{fig:NLOdist} but this time they are evaluated to NLO+NLL accuracy. The scale uncertainty bands in Figure~\ref{fig:NLONLLdist} are slightly narrower than the ones at NLO (shown in Figure~\ref{fig:NLOdist}) as a consequence of the inclusion of resummed soft gluon emission corrections.
Figure~\ref{fig:NLONLLnorm} shows the distributions at NLO+NLL normalized to their total cross section, i.e. the height of each bin is divided by the total cross section. Consequently, the sum of the height of all bins in each distribution (including the ones outside the range shown in the figure) is equal to 1. The scale uncertainty bands become extremely thin. For this reason we decided to show only the central value ($\mu_f = M/2, \, \mu_h = M, \, \mu_s = M/\bar{N}$) for both the scalar and the pseudoscalar case.  By examining Figure~\ref{fig:NLONLLnorm} one sees that, with the exception of the $p_T^t$ differential distribution, the shape of the normalized distributions in the scalar and pseudo-scalar cases is quite different. In the $M, M_{t \bar{t}}$ and $p_T^X$ distributions, the peak of the scalar coupling distribution is more pronounced and shifted to the left with respect to the peak of the pseudoscalar coupling distribution. As mentioned in the Introduction, the shape difference between distributions with scalar and pesudoscalar couplings can be experimentally used to probe the pseudoscalar component of the Yukawa coupling.

\begin{figure*}
	\begin{center}
		\begin{tabular}{cc}
			\includegraphics[width=7.cm]{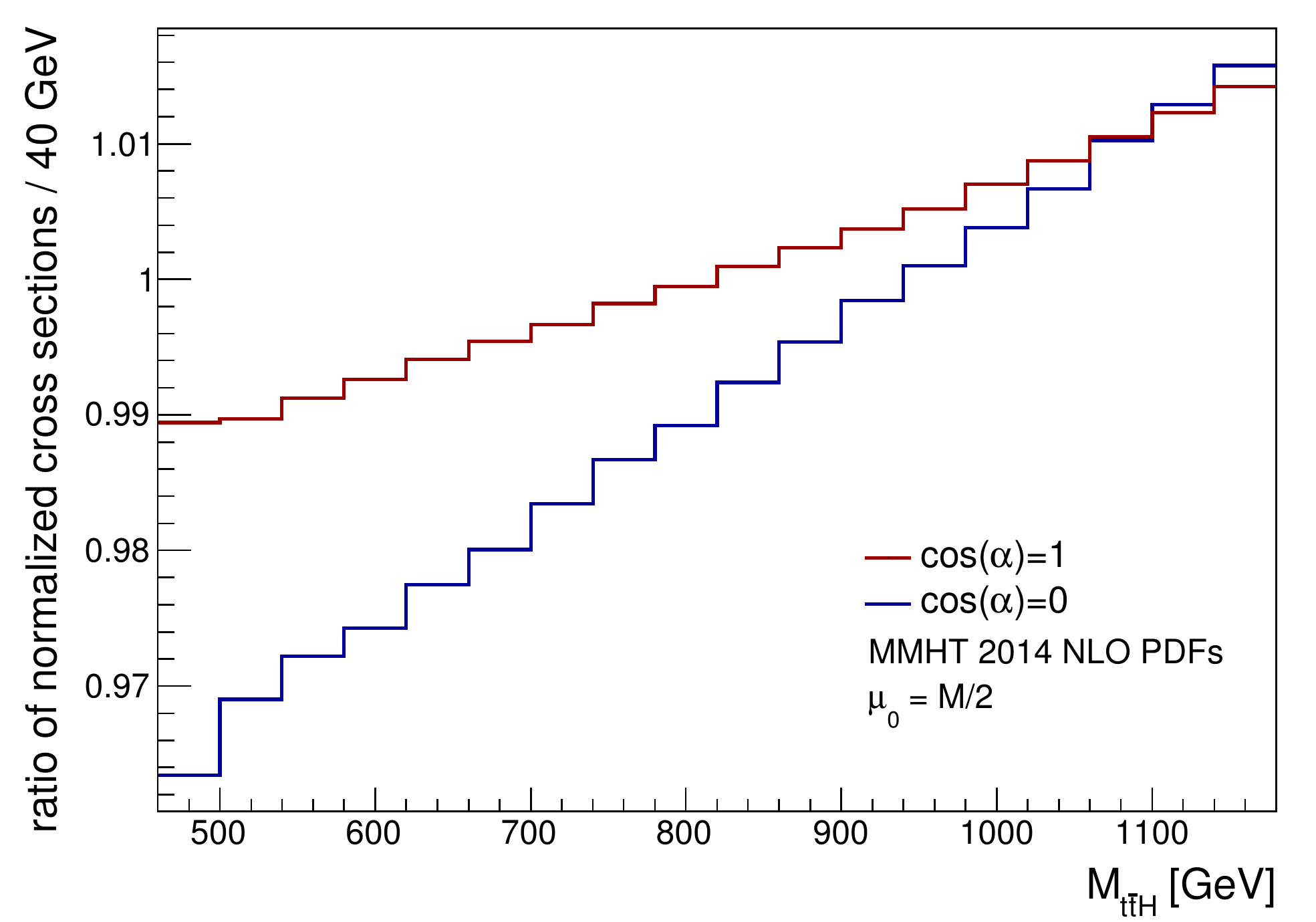} & \includegraphics[width=7.cm]{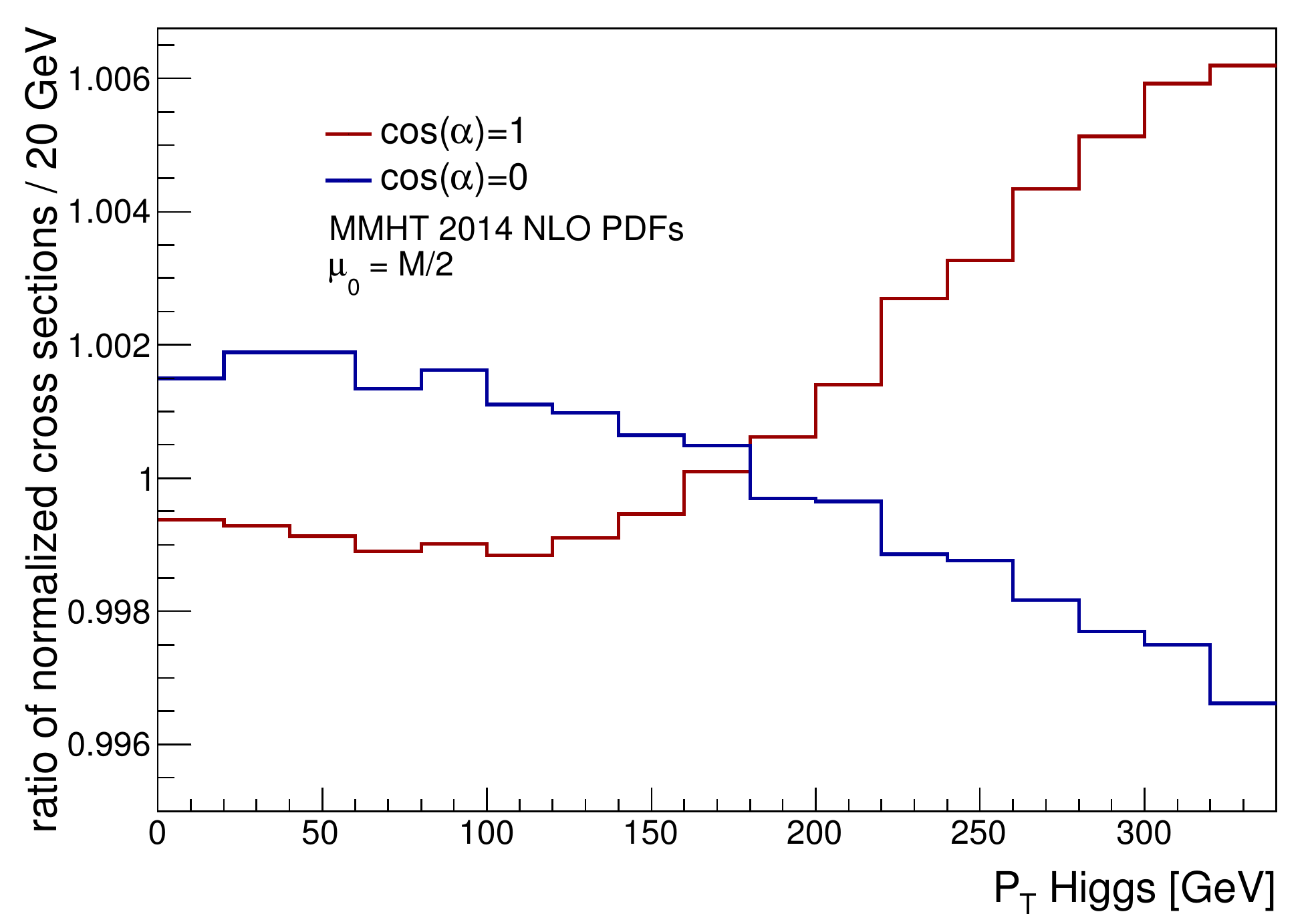} \\
			\includegraphics[width=7.cm]{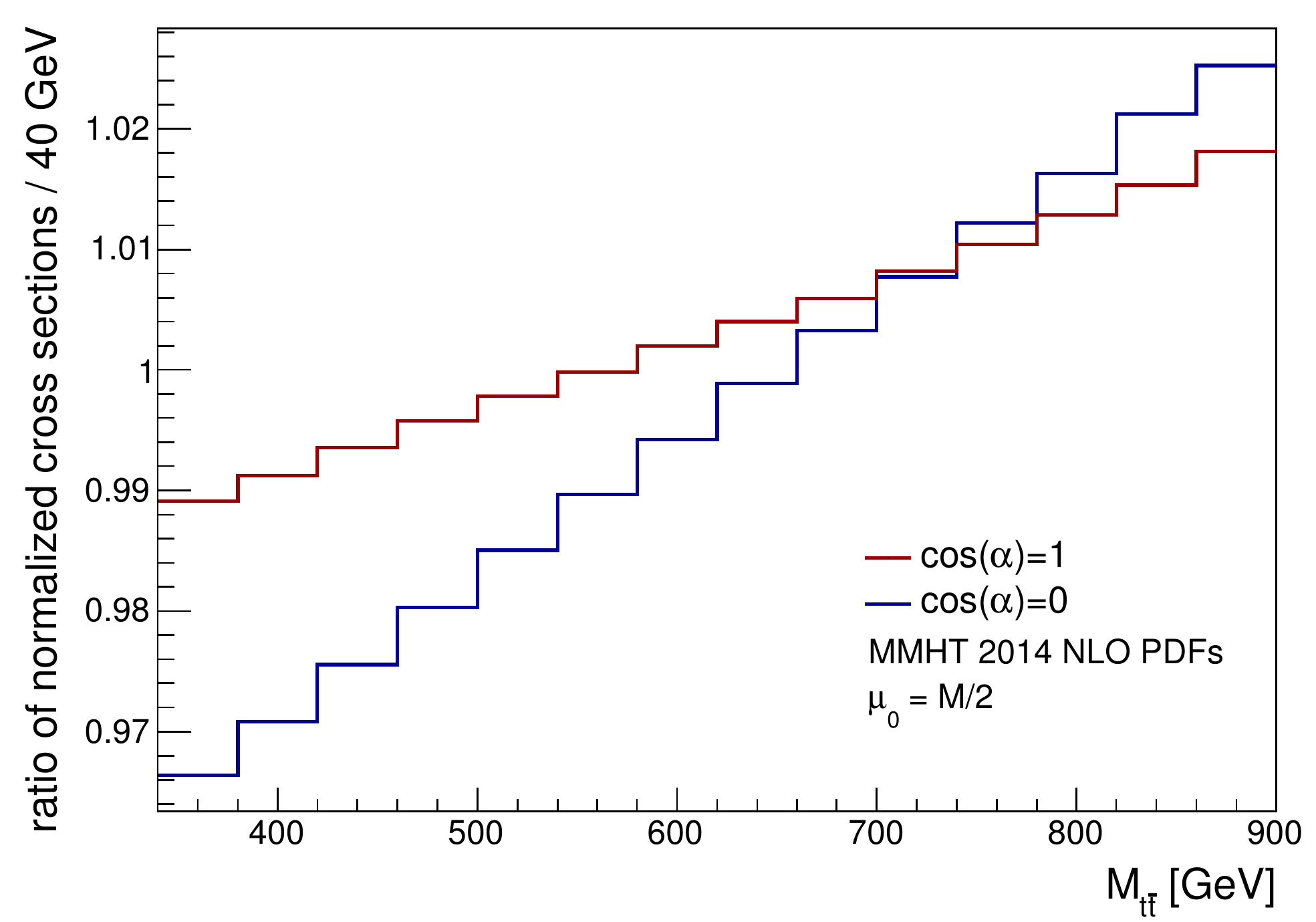} & \includegraphics[width=7.cm]{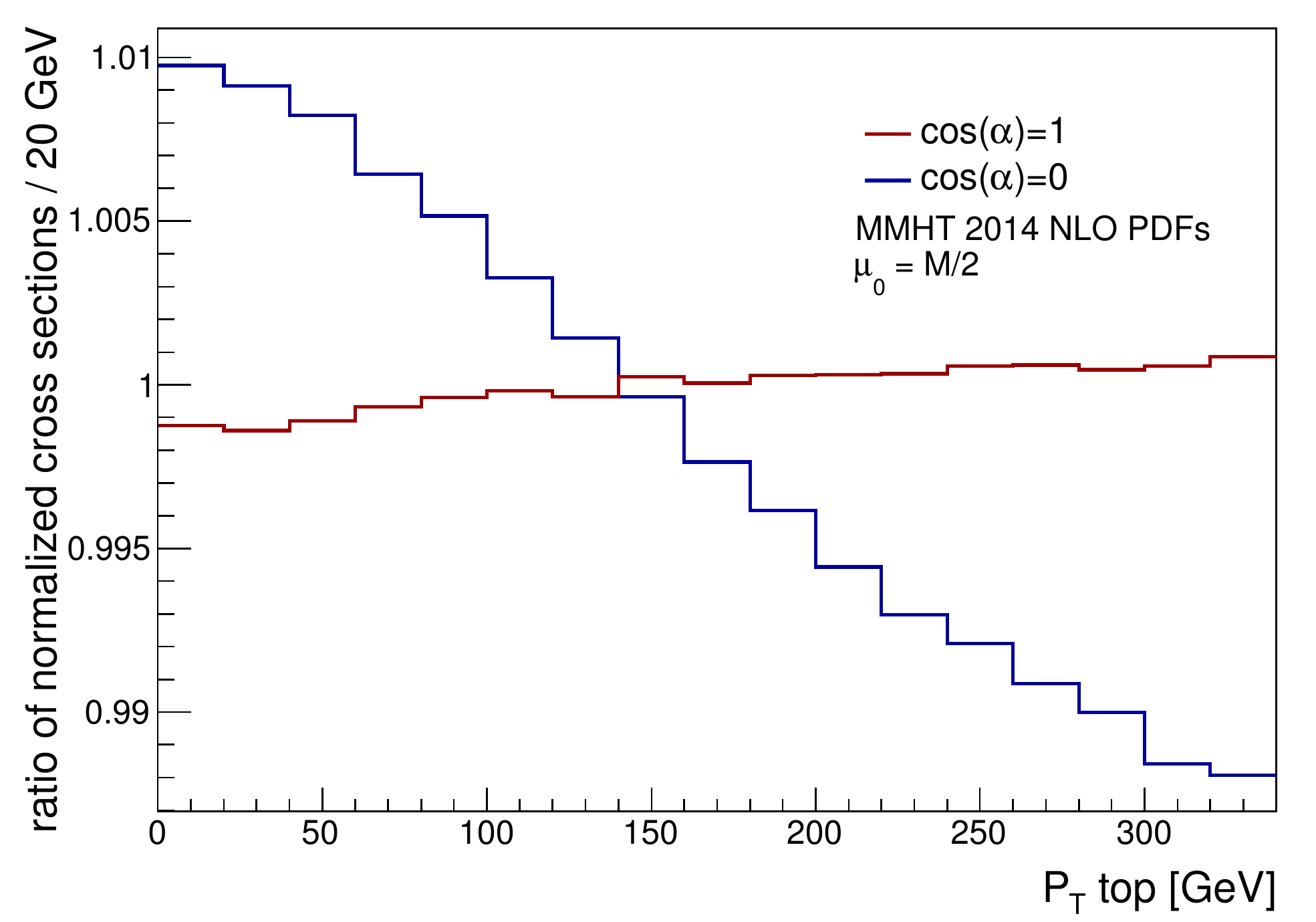}
		\end{tabular}
	\end{center}
    \caption{Ratio of the normalized distributions at NLO+NLL over NLO distributions. \label{fig:ratios}} 
\end{figure*}

One might wonder if there is a difference in shape between the fixed order and the matched distributions. It turns out that the NLO normalized distributions are extremely similar to the ones in Figure~\ref{fig:NLONLLnorm} because the resummed soft emission corrections have a very small impact on the shape of the distributions. This can be seen in Figure~\ref{fig:ratios}, which shows the ratio of the normalized distributions at NLO+NLL over the normalized distributions evaluated to NLO. The relative difference between these distributions is consistently smaller than a few percent.

\begin{figure*}
	\begin{center}
		\begin{tabular}{cc}
            			\includegraphics[width=7.cm]{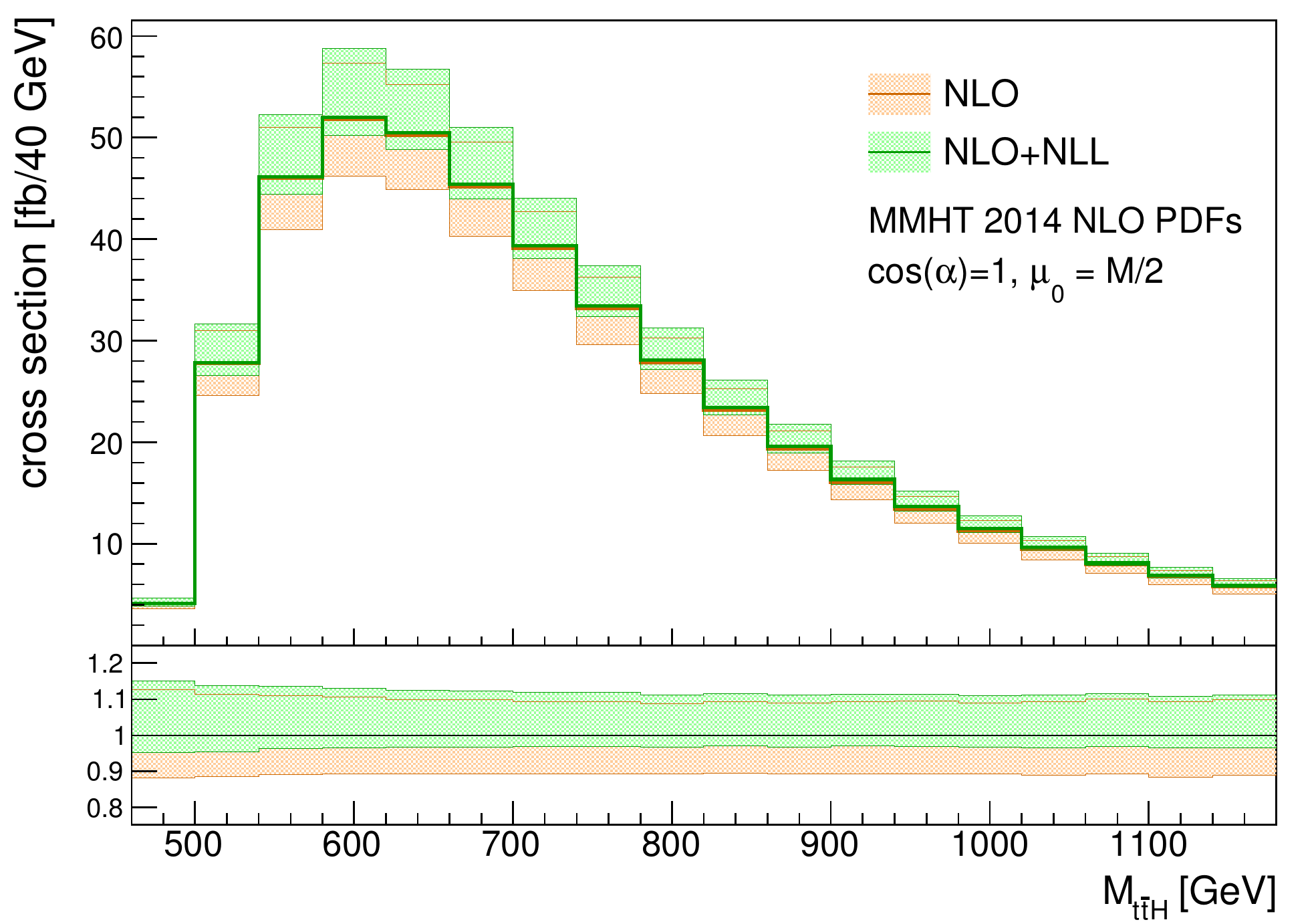} & \includegraphics[width=7.cm]{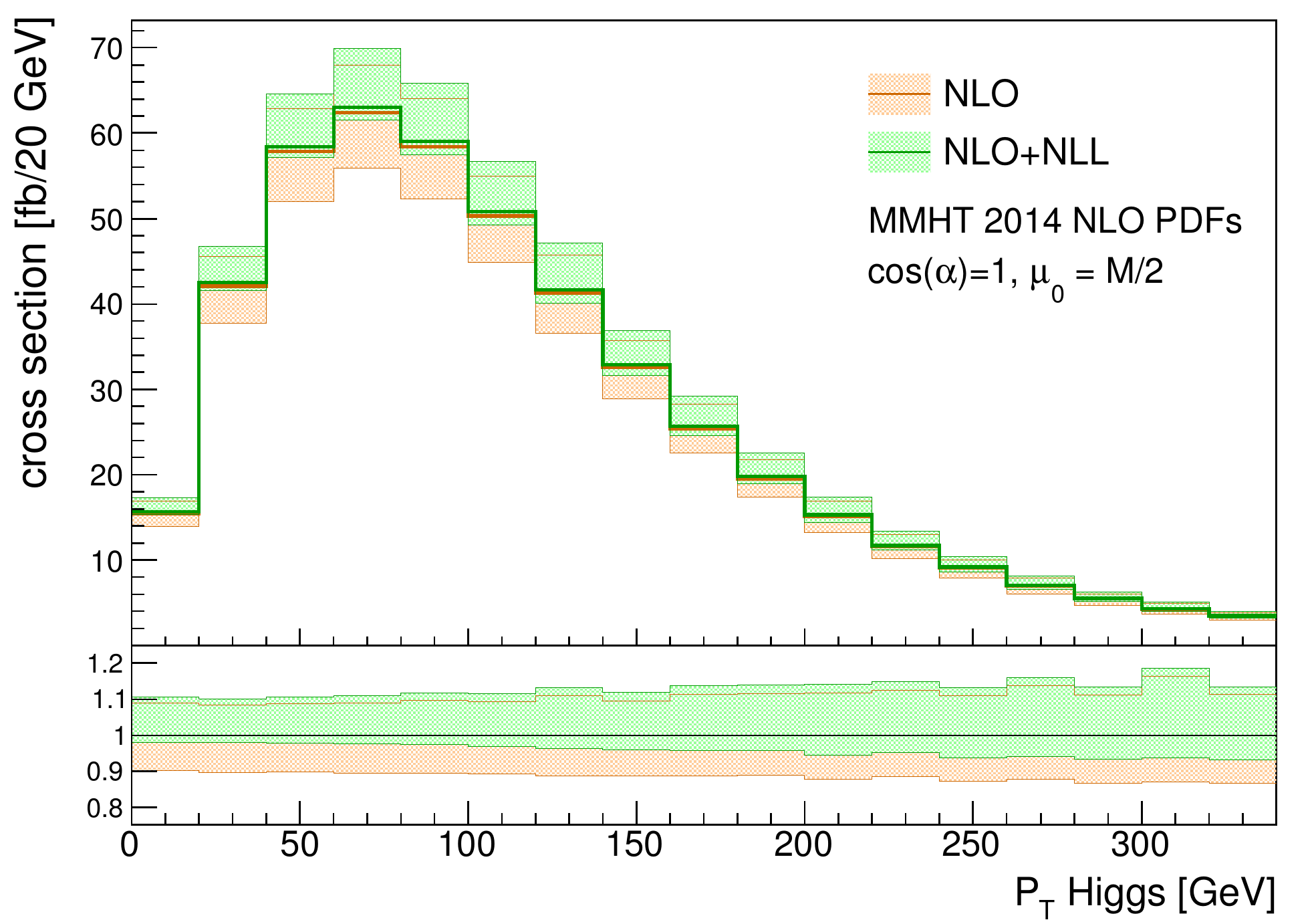} \\
			\includegraphics[width=7.cm]{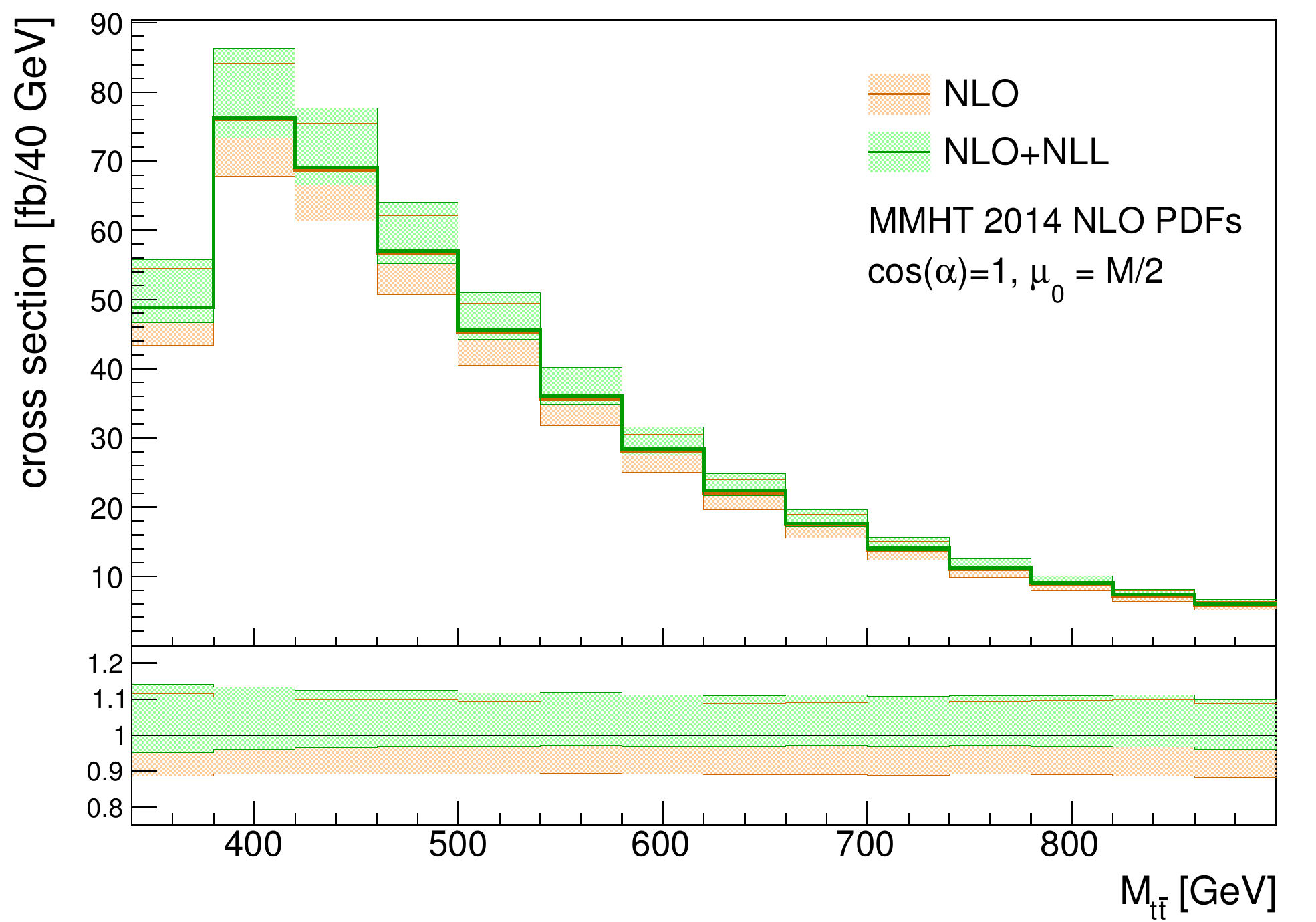} & \includegraphics[width=7.cm]{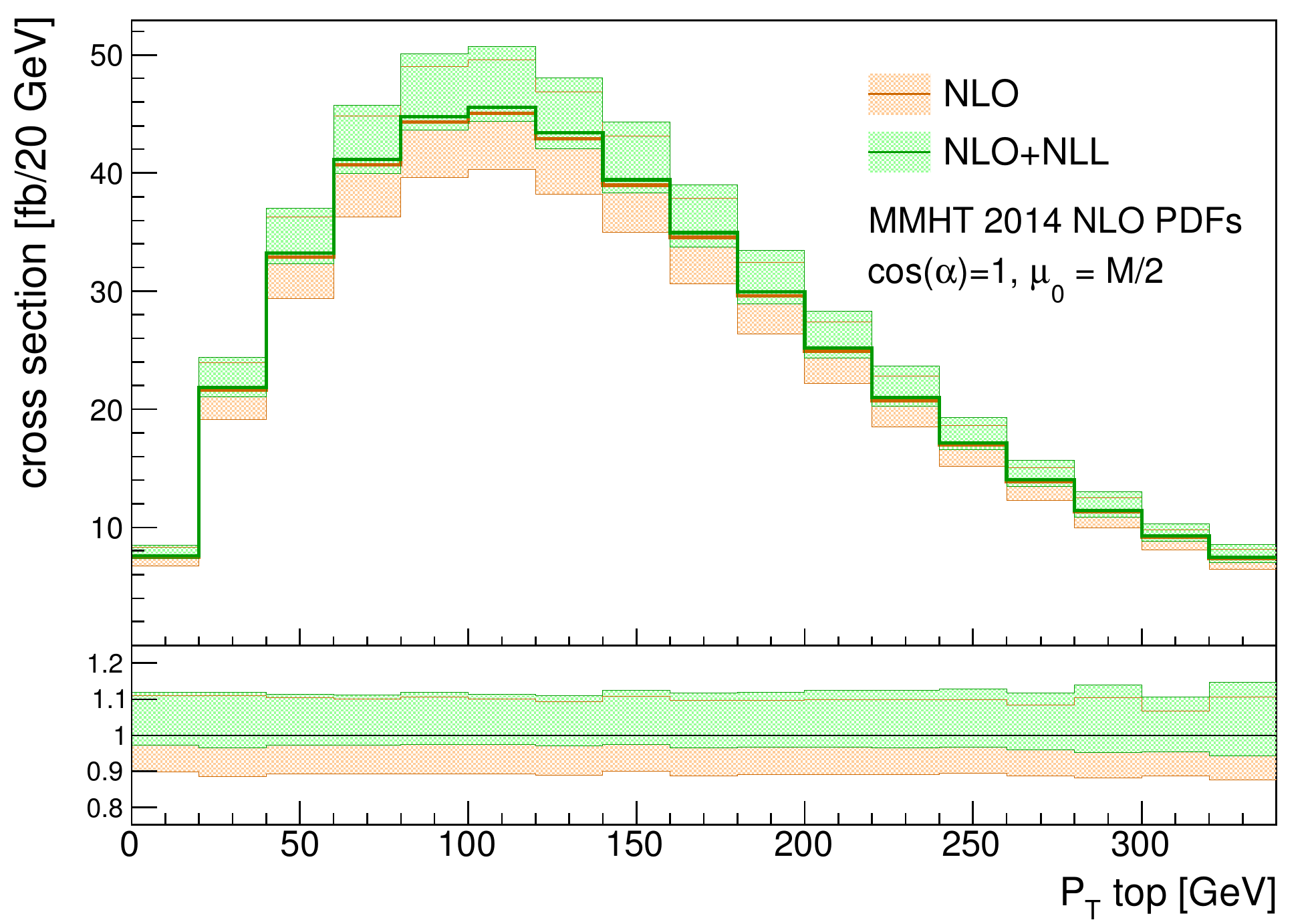}

		\end{tabular}
	\end{center}
    \caption{Comparison between NLO and NLO+NLL distributions for $\alpha = 0$. \label{fig:comp0}}   
\end{figure*}

\begin{figure*}
	\begin{center}
		\begin{tabular}{cc}
        \includegraphics[width=7.cm]{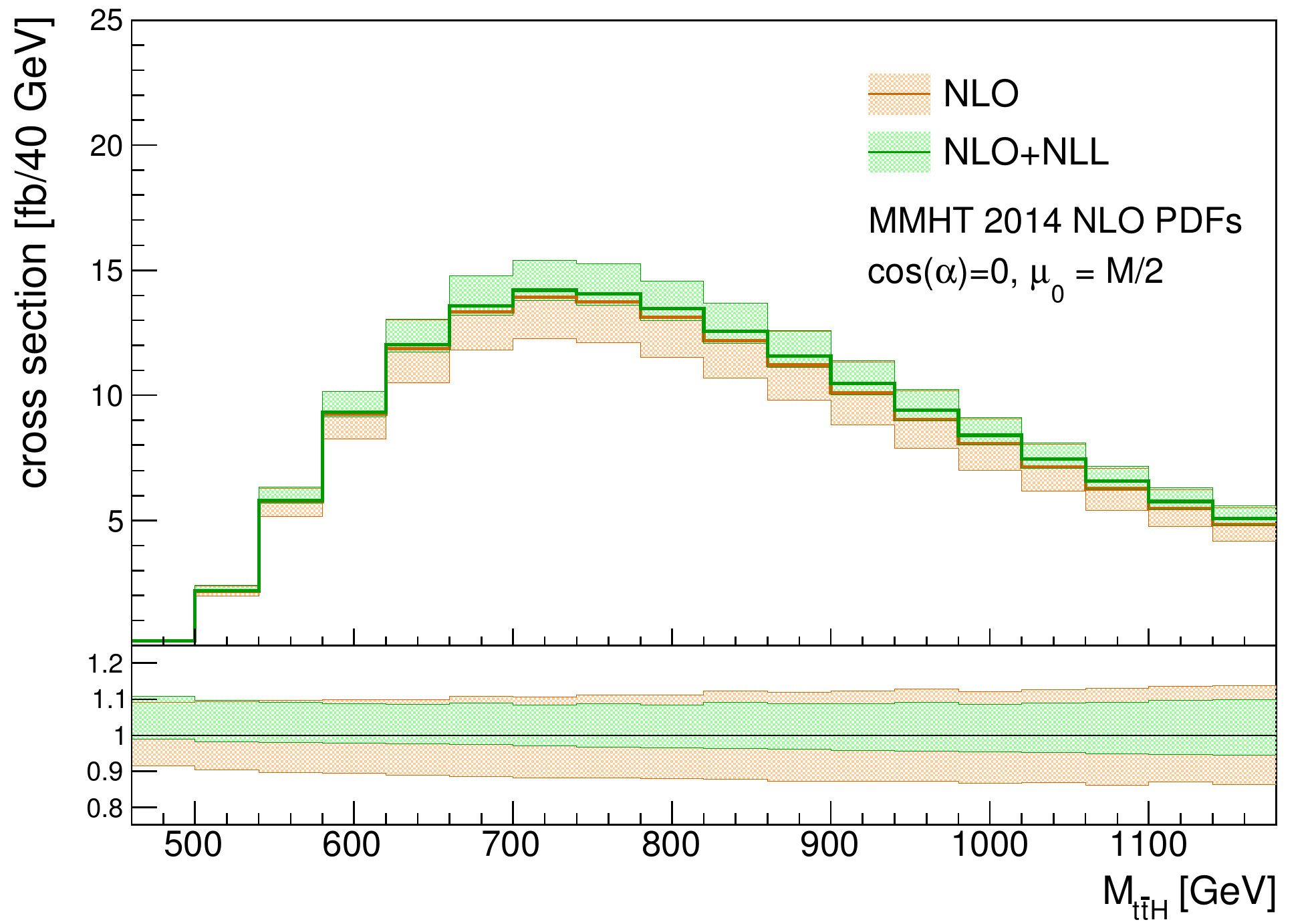} & \includegraphics[width=7.cm]{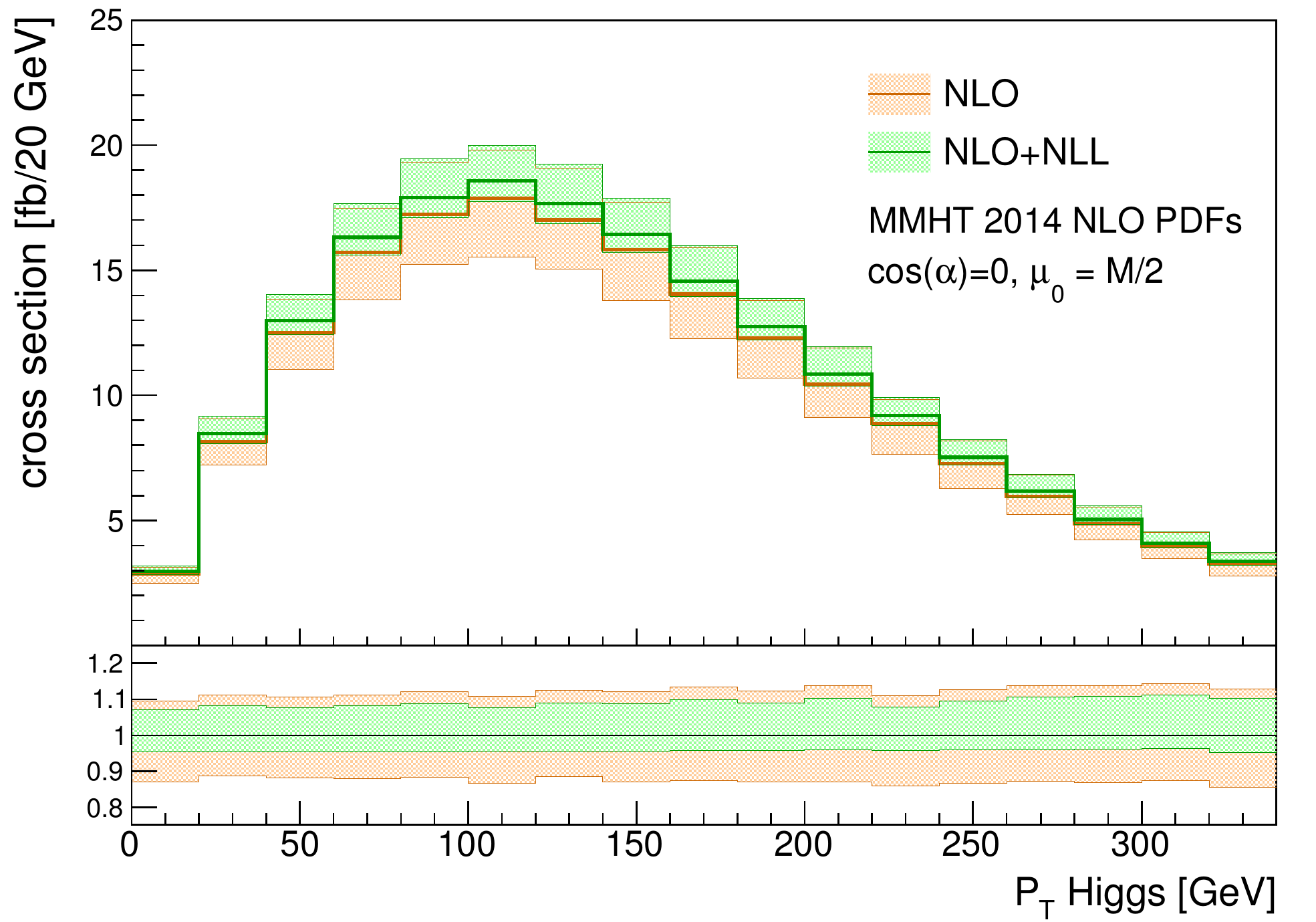} \\
			\includegraphics[width=7.cm]{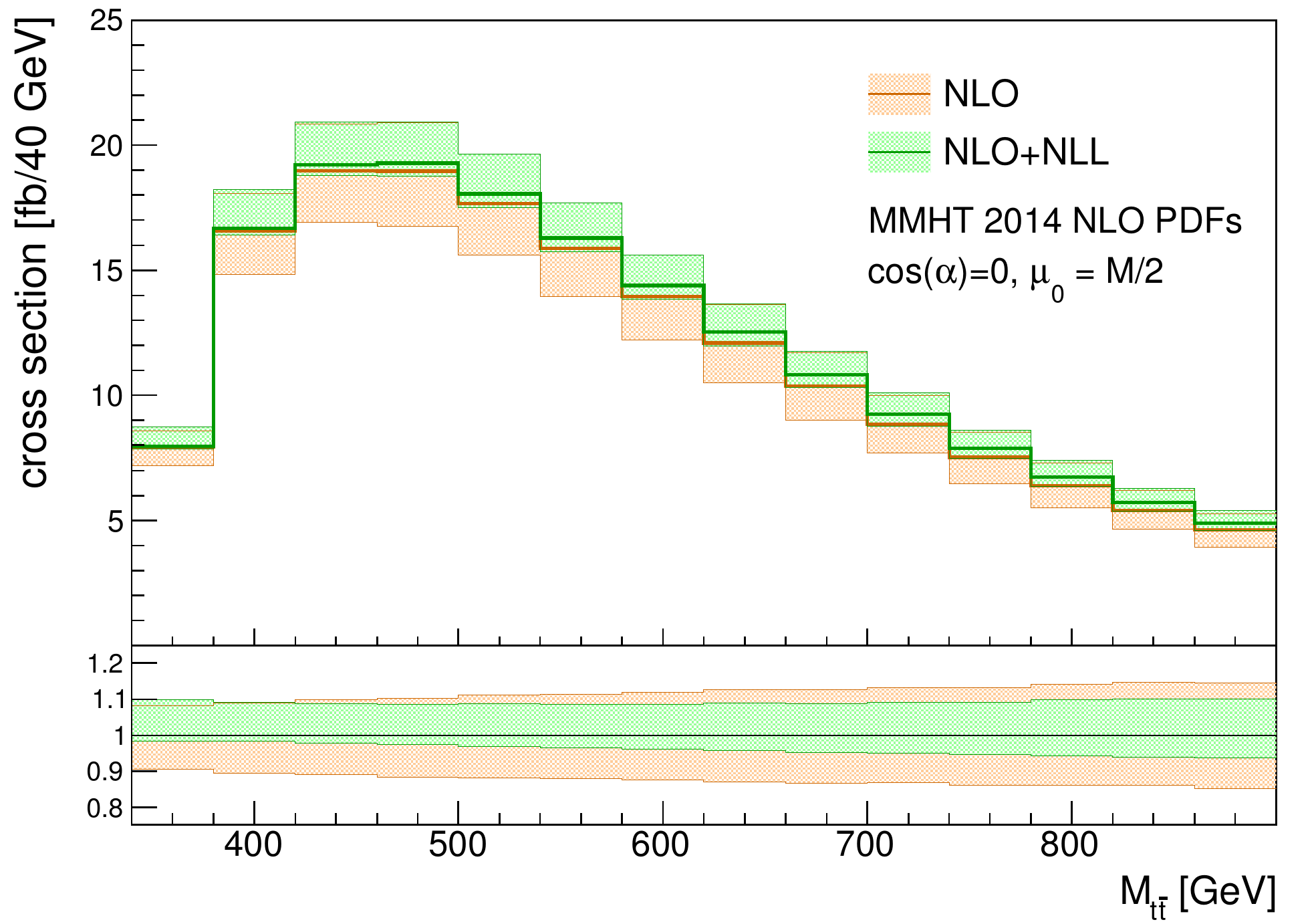} & \includegraphics[width=7.cm]{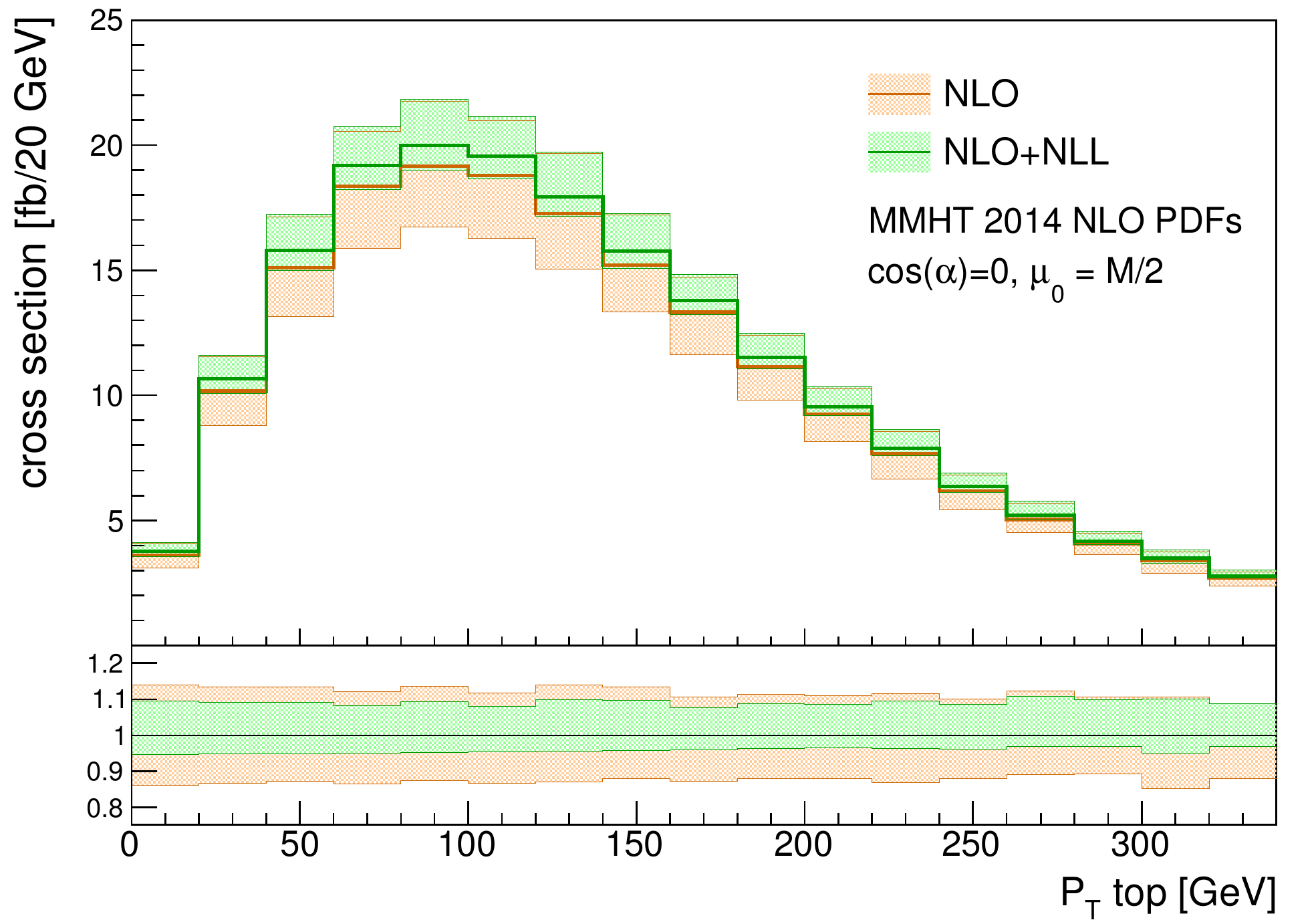} 
		\end{tabular}
	\end{center}
    \caption{Comparison between NLO and NLO+NLL distributions for $\alpha = \pi/2$. \label{fig:comppio2}}   
\end{figure*}

Figures~\ref{fig:comp0} and~\ref{fig:comppio2} show the comparison between the distribution evaluated to NLO (orange band) and to NLO+NLL (green band). Figure~\ref{fig:comp0} corresponds to the pure SM Higgs scenario while Figure~\ref{fig:comppio2} refers to the pure pseudoscalar case. The NLO+NLL bands are consistently overlapped with the upper part of the corresponding NLO bands. This behavior is similar to the one observed for the associated production of a top quark pair with a $W$ boson~\cite{Broggio:2016zgg} and the associated production of a top quark pair with a $Z$ boson~\cite{Broggio:2017kzi}.

\section{Experimental Motivation}
\label{sec:expan}

The cross-section for the $t\bar{t}X_0$ production channel has a very simple dependence on the mixing angle. This implies that the calculation of the cross section in the pure scalar case, $\sigma_H$, and in the pure pseudoscalar case, $\sigma_A$, allows us to fully reconstruct the dependence for any arbitrary $\alpha$. Therefore, the measurement of the total cross section by itself allows one to probe the pseudoscalar component of the top quark Yukawa coupling. This dependence on the mixing angle is also true for the aforementioned differential cross section distributions in any particular bin. As a result, the expected uncertainty on $|\cos \alpha|$ can be easily calculated as a function of the cross section.
In the SM scenario, assuming an hypothetical cross section measurement of $\sigma=\sigma_H\pm \Delta\sigma$, one finds
\begin{eqnarray}
\Delta|\cos \alpha| & = & \frac{\partial |\cos \alpha| }{\partial \sigma} \Delta \sigma \nonumber \\ 
& = & \frac{1}{2} \frac{1}{\sigma_H-\sigma_A} \Delta\sigma  \nonumber \\ 
& \approx & {0.86^{+0.09}_{-0.03} } \left ( {\frac{\sigma_H}{\Delta \sigma}} \right )^{-1} \, ,
\label{eq:}
\end{eqnarray}
according to Table~\ref{tab:CSHp13hM}, at NLO+NLL. The uncertainty on $|\cos \alpha|$ is inversely proportional to the statistical significance of the production channel. The graphical representation of the lower limit on $\cos \alpha$ at 68.3\% CL with the statistical significance of the SM Higgs boson associated production with a top quark pair is presented in Figure~\ref{fig:limit}. An eventual five sigma observation of this channel, in the SM hypothesis, would set a limit of \mbox{$|\cos \alpha| > 0.83_{-0.02}^{+0.01}$} at 68.3\% CL, growing asymptotically to one with the increase of significance. The error on this limit, represented by the blue shaded band in Figure~\ref{fig:limit}, emerges from the uncertainty on the cross-sections in Table~\ref{tab:CSHp13hM} due to scales variations.

\begin{figure*}
	\begin{center}
		\begin{tabular}{cc}
			\includegraphics[width=7.cm]{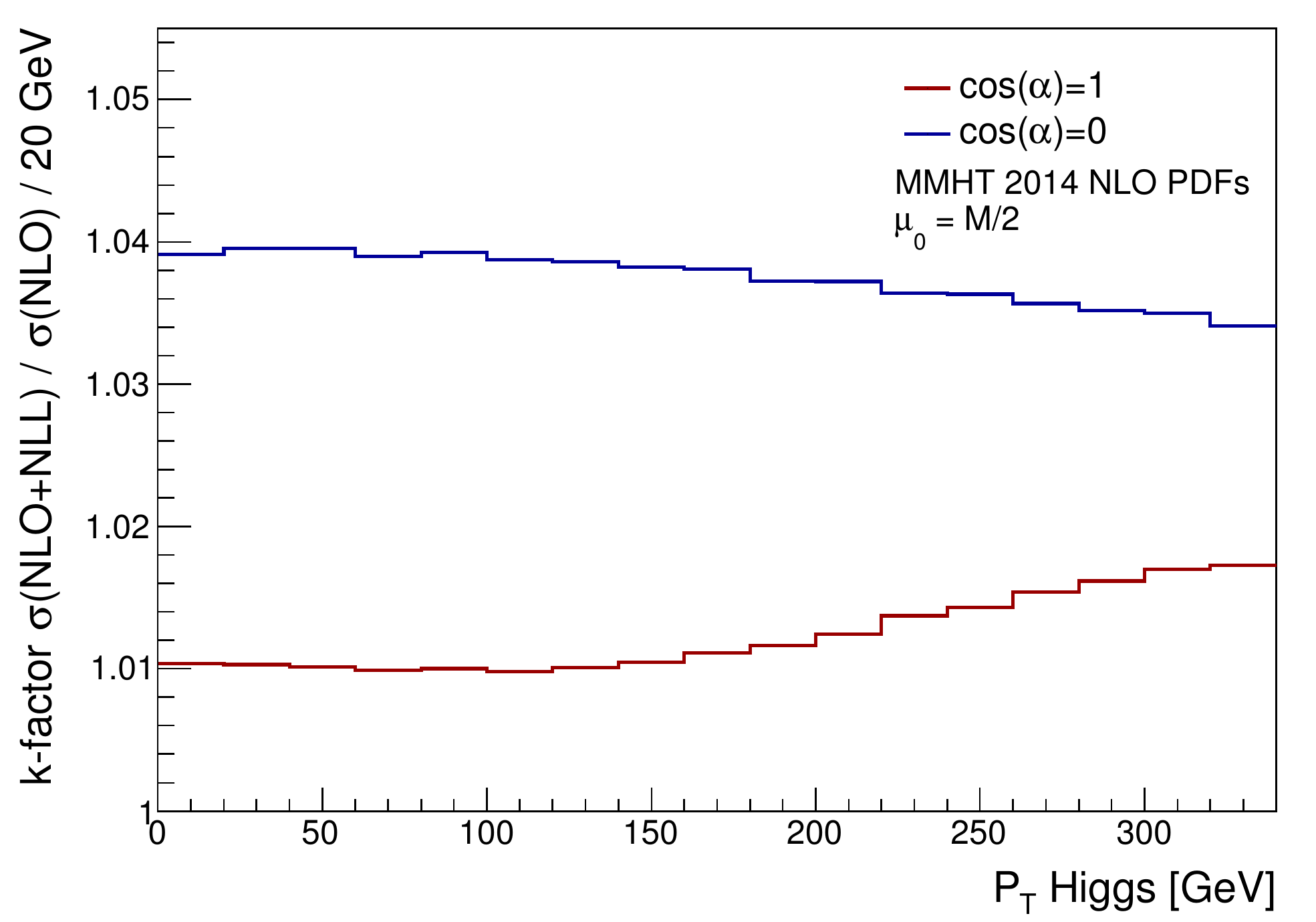} & \includegraphics[width=7.cm]{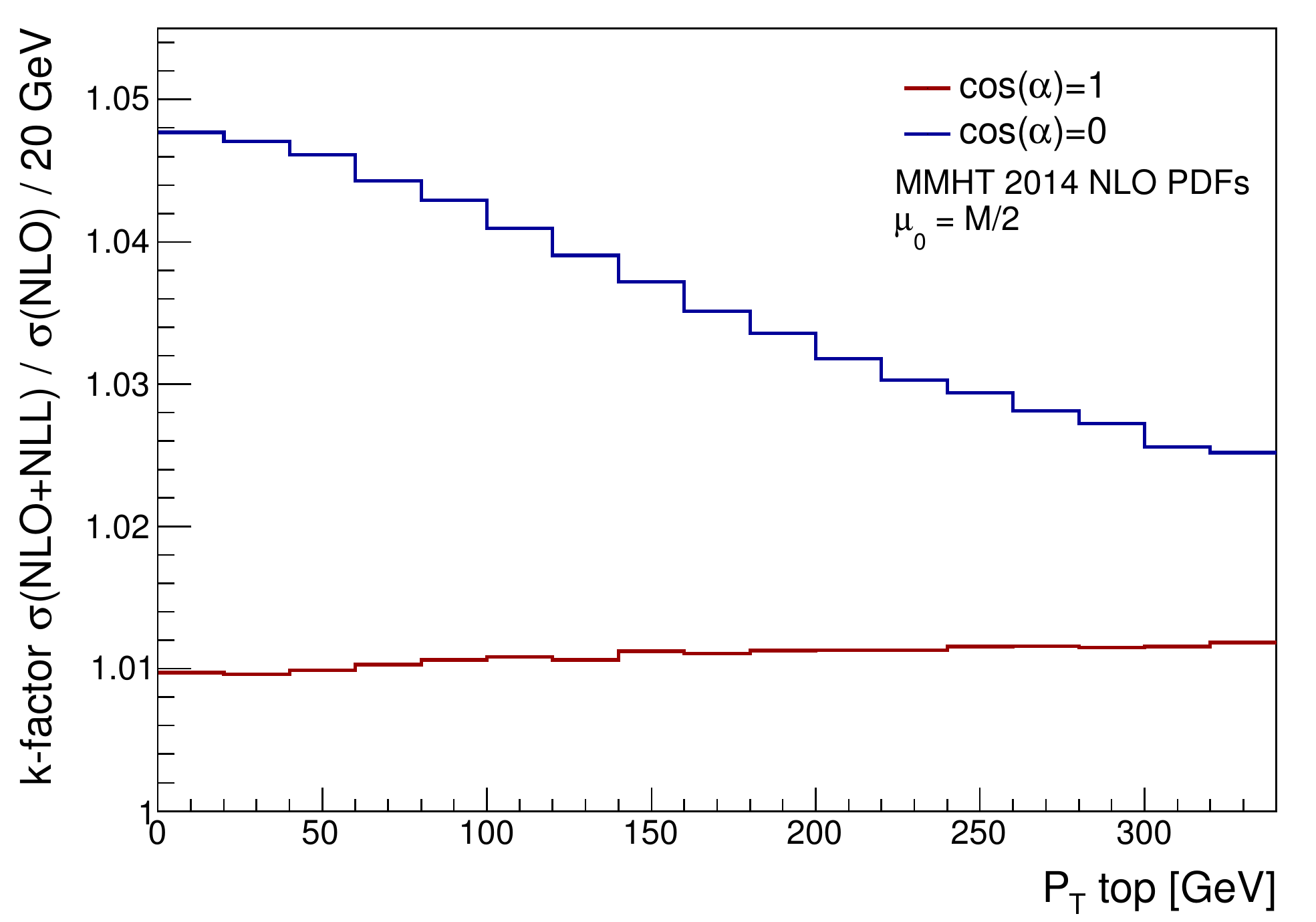}
		\end{tabular}
	\end{center}
    \caption{Ratio of the cross-section at NLO+NLL over NLO distributions ($k$-factors). \label{fig:ratiosxsec}} 
\end{figure*}

\begin{figure*}
\begin{center}
\includegraphics[width=12.2cm]{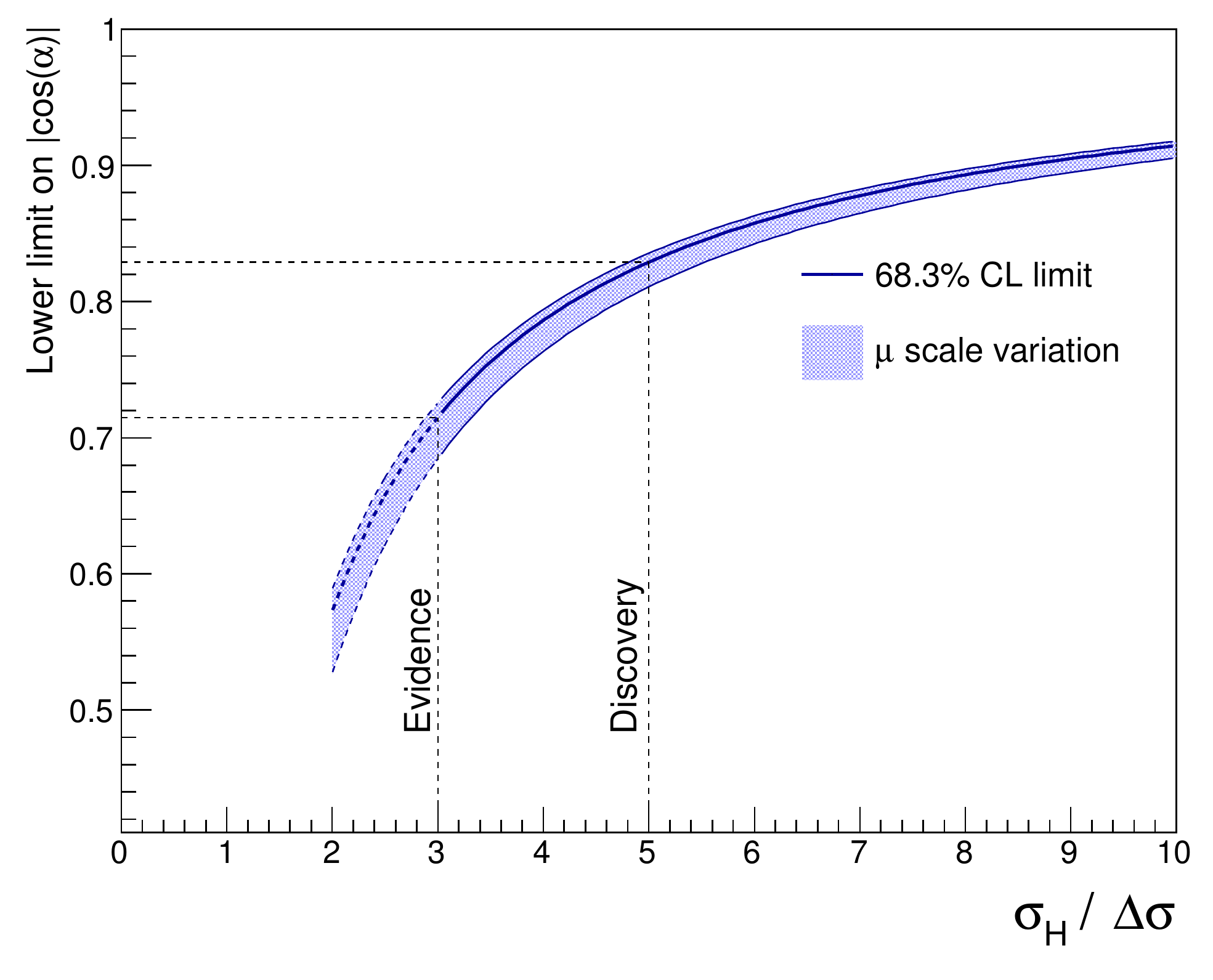}
\caption{Dependence of the lower limit on $|\cos(\alpha)|$ at 68.3\% CL with the statistical significance of the SM Higgs boson associated production with a top quark pair, at NLO+NLL. The blue error band around the central limit line represents the scale uncertainty.}
\label{fig:limit}
\end{center}
\end{figure*}

The theoretical uncertainty on the $t\bar{t}X$ signal cross section, in particular the scale dependence, comprises one of the components of the full systematic error taken into account in experimental measurements at the LHC. The observed improvement on the scale uncertainty at NLO+NLL is expected to significantly decrease the impact of the scale dependences on the Monte Carlo signal estimation, and therefore, on the overall systematic error~\cite{CMS:2016zbb}. Moreover, the $k$-factors associated with the ratio of the NLO+NLL cross section with respect to the NLO calculation,
\begin{equation}
k = \sigma^{\textrm{NLO+NLL}}/\sigma^{\textrm{NLO}} \, ,
\end{equation}
range from 1.010 to 1.012 (1.048 to 1.025) for the transverse momentum of the top quark in the pure scalar (pseudoscalar) scenario, as can be seen in Figure~\ref{fig:ratiosxsec}. This information is useful to reweight the simulated transverse momenta of the top quark and $X_0$ boson in the experimental measurement analyses, for any specific mixture of the two components.

While the cross section measurement does not require an experimental full event kinematic reconstruction, the implementation of the differential distributions evaluated  in Section~\ref{sec:numbers}, and eventual new angular asymmetries, naturally demands complete kinematic information about the top quarks and the $X_0$ boson. The different behavior of these observables for the scalar and pseudoscalar cases are expected to significantly improve the sensitivity to the mixing angle $\alpha$, when combined with the cross section result. In particular, the two dimensional distribution between the top-$X_0$ and the anti-top-$X_0$ angles in the $t\bar{t}X_0$ rest frame, recently presented in~\cite{AmorDosSantos:2017ayi}, is a potential  candidate to provide an angular asymmetry which would allow to probe the top quark Yukawa coupling with further precision.

The theoretical understanding of the shapes of the transverse momentum and invariant mass distributions is also extremely important for the full kinematic reconstruction of $t \bar{t} X_0$ dilepton events. The event reconstruction is based on six kinematic equations with quadratic dependences, and therefore, gives rise to more than one possible solution~\cite{Aad:2012ky,Santos:2015dja}. As such, these variables can be used as inputs to calculate the likelihood of a given solution to be consistent with the event. For example, the reconstruction of the $t\bar{t}X_0$ system performed in \cite{AmorDosSantos:2017ayi}, with two opposite charged leptons in the final state, makes use of a likelihood function calculated as the product of one-dimensional probability density functions (p.d.f.). These p.d.f.s are built from $p_T$ distributions of the neutrino, anti-neutrino, top quark, anti-top quark, and $t\bar{t}$ system, respectively $P({p_T}_{\nu})$, $P({p_T}_{\bar{\nu}})$, $P({p_T}_{t})$, $P({p_T}_{\bar{t}})$ 
and $P({p_T}_{t\bar{t}})$, at parton level:
\begin{eqnarray}
\label{equ:a7}
\ensuremath{L_{t\bar{t}X_0} ~~&\sim&~~ \frac{1}{{p_T}_{\nu} {p_T}_{\bar{\nu}}} P({p_T}_{\nu}) P({p_T}_{\bar{\nu}})  P({p_T}_{t}) P({p_T}_{\bar{t}}) \nonumber \\ 
& \times &  P({p_T}_{t\bar{t}}) P(m_t,m_{\bar{t}}) P(m_{X_0})} \, .
\end{eqnarray}
The two-dimensional p.d.f. of the top quark masses, $P(m_t,m_{\bar{t}})$, and the one-dimensional p.d.f. of the 
$X_0$ candidate mass, $P(m_{X_0})$, are also included. In addition, the kinematic distributions presented in this paper might also be used as input variables in similar multivariate methods to determine the right pairing of jets in the reconstruction process. These distributions can also be used to repair the distortion caused by the detector response, kinematic cuts applied in the trigger and in the offline event reconstruction and selection.

\section{Conclusions}
\label{sec:conclusions}

In this work we studied the associated production of a top-antitop pair and a scalar $X_0$ which is allowed to couple to the top quark current with a mixture of scalar and pseudoscalar couplings. The scalar coupling is the SM Yukawa coupling modulated by a factor $\cos \alpha$, where $\alpha$ is the angle which parameterizes the relative strength of the scalar and pseudoscalar components of the coupling. We evaluated the total cross section for this process for the special cases $\alpha = 0$ (SM Higgs) and $\alpha = \pi/2$ (pure pseudoscalar coupling) to NLO+NLL accuracy, where the NLL resummation was carried out in the soft gluon emission limit. We showed that it is straightforward to use this information in order to evaluate the total cross section for arbitrary values of $\alpha$.

Recent studies point to the possibility of a reliable  kinematic reconstruction of the top, antitop and $X_0$~momenta. For this reason, we studied differential distributions with respect to the final state invariant mass, $X_0$ transverse momentum, top-pair invariant mass and top-quark transverse momentum to NLO+NLL accuracy. By comparing the 
$\alpha = 0$ and $\alpha = \pi/2$ cases, we showed that the shapes of the first three distributions listed above depend quite strongly on the scalar or pseudoscalar nature of the coupling between the top quark and the Higgs boson. In addition, these shapes are stable with respect to the inclusion of higher-order soft emission corrections. Results for the distributions at arbitrary values of $\alpha$ can be easily obtained by considering an appropriate bin-by-bin combination of the results at $\alpha = 0$ and $\alpha = \pi/2$.

It is shown that a possible future observation of the associated production of a top quark pair with a Higgs boson in the SM hypothesis could immediately establish a limit on the pseudoscalar component of the top quark Yukawa coupling of \mbox{$|\cos \alpha| > 0.83_{-0.02}^{+0.01}$}, at 68.3\% CL.  Moreover, a full kinematic reconstruction of the final state momenta for this process could provide valuable information by means of angular distributions and asymmetries, which can be combined with the cross section result. The theoretical understanding of the aforementioned differential kinematic distributions is also extremely useful for the experimental measurement itself, as they can provide $k$-factors for transverse momentum reweighting, and probability density functions for neutrino reconstruction in dilepton events.

\section*{Acknowledgments}
We thank R.~Frederix, G.~Ossola and N.~Castro for useful discussions and for reading the manuscript. The in-house Monte Carlo code which we developed and employed to evaluate the (differential) cross sections presented in this paper was run on the computer cluster of the Center for Theoretical Physics at the Physics Department of New York City College of Technology. This work was supported by the CUNY	Summer Collaborative Research Opportunity Grant Award 80232-20~17 and PSC-CUNY Awards 60061-00~48 and 60185-00~48. The work of A.F. is supported in part by the National Science Foundation under Grant No. PHY-1417354.


%
%

\end{document}